\documentclass[twocolumn,preprintnumbers,prx,aps,amssymb,amsmath,showpacs,floatfix,superscriptaddress,longbibliography]{revtex4-1}

\usepackage{graphicx}
\usepackage{epsfig}
\usepackage{natbib}
\usepackage{dcolumn}
\usepackage{bm}
\usepackage{soul}
\usepackage{sidecap}
\usepackage[usenames,dvipsnames]{color}
\usepackage{setspace}
\usepackage[colorlinks=true,linkcolor=blue,urlcolor=blue,citecolor=blue]{hyperref}
\usepackage[usenames,dvipsnames,svgnames,table]{xcolor}
\usepackage{wasysym}
\usepackage{ulem}
\usepackage{multirow}

\begin{document}
%\preprint{1p0mt}

\newcommand{\Nd}{La$_{1.6-x}$Nd$_{0.4}$Sr$_{x}$CuO$_4$}
\newcommand{\Eu}{La$_{1.8-x}$Eu$_{0.2}$Sr$_{x}$CuO$_4$}
\newcommand{\LSCO}{La$_{2-x}$Sr$_{x}$CuO$_4$}
\newcommand{\YBCO}{YBa$_{2}$Cu$_{3}$O$_y$}
\newcommand{\PCCO}{Pr$_{2-x}$Ce$_{x}$CuO$_4$}
\newcommand{\BiONE}{Bi$_2$La$_{2-x}$Sr$_{x}$CuO$_{6 + \delta}$}
\newcommand{\BiTWO}{Bi$_2$Sr$_2$CaCuO$_{8 + x}$}

\newcommand{\ie}{{\it i.e.}}
\newcommand{\eg}{{\it e.g.}}
\newcommand{\etal}{{\it et al.}}

\newcommand{\TN}{$T_{\rm N}$}
\newcommand{\Tc}{$T_{\rm c}$}
\newcommand{\Tstar}{$T^\star$}
\newcommand{\Tsdw}{$T_{\rm SDW}$}

\newcommand{\Hc}{$H_{\rm c2}$}

\newcommand{\pstar}{$p^\star$}
\newcommand{\pcdw}{$p_{\rm CDW}$}
\newcommand{\po}{$p_0$}

\newcommand{\RH}{$R_{\rm H}$}
\newcommand{\nH}{$n_{\rm H}$}
\newcommand{\nrho}{$n_{\rho}$}

\newcommand{\Kzero}{$\kappa_0/T$}
\newcommand{\units}{$\mu \text{W}/\text{K}^2\text{cm}$}
\newcommand{\p}[1]{\left( #1 \right)}
\newcommand{\Dd}[2]{\frac{\text{d} #1}{\text{d}#2}}

%%%%%%%%%%%%%%%%%%%%%%%%%%%% TITLE

\title{
Fermi-surface transformation across the pseudogap critical point 
of the cuprate superconductor \Nd
}

%%%%%%%%%%%%%%%%%%%%%%%%%%%% AUTHORS

\author{C.~Collignon}
\email[]{clement.collignon@usherbrooke.ca}
\affiliation{D\'epartement de Physique and RQMP, Universit\'e de Sherbrooke, Sherbrooke, Qu\'ebec, Canada J1K 2R1}
\affiliation{Laboratoire de Physique et d'\'Etude des Mat\'eriaux, 
\'Ecole Sup\'erieure de Physique et de Chimie Industrielles (CNRS), Paris 75005, France}

\author{S.~Badoux}
\affiliation{D\'epartement de Physique and RQMP, Universit\'e de Sherbrooke, Sherbrooke, Qu\'ebec, Canada J1K 2R1}

\author{S.~A.~A.~Afshar} 
\affiliation{D\'epartement de Physique and RQMP, Universit\'e de Sherbrooke, Sherbrooke, Qu\'ebec, Canada J1K 2R1}

\author{B.~Michon}
\affiliation{D\'epartement de Physique and RQMP, Universit\'e de Sherbrooke, Sherbrooke, Qu\'ebec, Canada J1K 2R1}

\author{F.~Lalibert\'e}
\affiliation{D\'epartement de Physique and RQMP, Universit\'e de Sherbrooke, Sherbrooke, Qu\'ebec, Canada J1K 2R1}

\author{O.~Cyr-Choini\`ere} 
\altaffiliation{Present address: Department of Physics, McGill University, Montreal, Qu\'{e}bec H3A 2T8, Canada}
\affiliation{D\'epartement de Physique and RQMP, Universit\'e de Sherbrooke, Sherbrooke, Qu\'ebec, Canada J1K 2R1}

\author{J.-S.~Zhou}
\affiliation{Texas Materials Institute, University of Texas at Austin, Austin, Texas 78712, USA}

\author{S.~Licciardello}
\affiliation{High Field Magnet Laboratory (HFML-EMFL) and Institute for Molecules and Materials,
Radboud University, 6525 ED Nijmegen, The Netherlands}

\author{S.~Wiedmann}
\affiliation{High Field Magnet Laboratory (HFML-EMFL) and Institute for Molecules and Materials,
Radboud University, 6525 ED Nijmegen, The Netherlands}

\author{N.~Doiron-Leyraud}
\affiliation{D\'epartement de Physique and RQMP, Universit\'e de Sherbrooke, Sherbrooke, Qu\'ebec, Canada J1K 2R1}

\author{Louis~Taillefer}
\email[]{louis.taillefer@usherbrooke.ca}
\affiliation{D\'epartement de Physique and RQMP, Universit\'e de Sherbrooke, Sherbrooke, Qu\'ebec, Canada, J1K 2R1}
\affiliation{Canadian Institute for Advanced Research, Toronto, Ontario, Canada M5G 1Z8}

\date{\today}

\begin{abstract}

%%%%%%%%%%%%%%%%%%%%%%%%%%%%   ABSTRACT

The electrical resistivity $\rho$ and Hall coefficient \RH~of the tetragonal single-layer cuprate \Nd~were measured in magnetic fields up to $H = 37.5$~T, 
large enough to access the normal state at $T \to 0$, for closely spaced dopings $p$
across the pseudogap critical point at \pstar~$= 0.23$.
Below \pstar, both coefficients exhibit an upturn at low temperature, which gets more pronounced with decreasing $p$. 
Taken together, these upturns show that the normal-state carrier density $n$ at $T = 0$ drops upon entering the pseudogap phase. 
Quantitatively, it goes from $n = 1 + p$ at $p = 0.24$ to $n = p$ at $p = 0.20$.
By contrast, the mobility does not change appreciably, as revealed by the magneto-resistance.
The transition has a width in doping and some internal structure, whereby \RH~responds more slowly than $\rho$ to the opening of the pseudogap. 
We attribute this difference to a Fermi surface that supports both holelike and electronlike carriers in the interval $0.2 < p <Ê$\pstar,
with compensating contributions to \RH. 
Our data are in excellent agreement with recent high-field data on \YBCO~and \LSCO.
The quantitative consistency across three different cuprates shows that a drop in carrier density from $1 + p$ to $p$ is
a universal signature of the pseudogap transition at $T=0$.
We discuss the implication of these findings for the nature of the pseudogap phase.

%%%%%%%%%%%%%%%%%%%%%%%%%%%%%%%%%%%%%%%%%%%%%%%%%%%%%%%%%%%%%%%%%%%%%%%%%

\end{abstract}

\pacs{74.25.Fy, 74.70.Dd}
%74.25.Fy Transport properties (electric and thermal conductivity, thermoelectric effects, etc.)
%74.20.Rp Pairing symmetries (other than s-wave)
%74.70.Dd Ternary, quaternary, and multinary compounds (including Chevrel phases, borocarbides, etc.)

\maketitle

%%%%%%%%%%%%%%%%%%%%%%%%%%%%      INTRODUCTION  

\section{Introduction}

After more than two decades, the pseudogap phase of cuprate superconductors remains an enigma, the subject of active debate. 
Most experimental studies so far have been carried out either at high temperature, above the onset of superconductivity at \Tc, 
where signatures are typically broad,
or at low temperature, inside the superconducting phase, where it is difficult to separate the pseudogap from the superconducting gap. 
Experiments of a third kind are called for: in the $T=0$ limit, without superconductivity [\onlinecite{Broun2008}].
This can be achieved by applying large magnetic fields to suppress superconductivity. 

Twenty years ago, Ando, Boebinger and co-workers pioneered this approach with measurements of the electric resistivity $\rho(T)$ in the cuprate \LSCO~(LSCO), 
using pulsed fields up to 61~T [\cite{Ando1995,Boebinger1996}]. 
They discovered an upturn in $\rho(T)$ at low $T$, for hole concentrations (dopings) below $p \simeq 0.16$. 
The mechanism responsible for what was called 
a ``metal-to-insulator crossover" has remained unclear until very recently [\onlinecite{Laliberte2016}]
(see below).
Later on, Hussey and co-workers showed that $\rho(T)$ in LSCO decreases linearly as $T \to 0$ at $p = 0.18$, and up to $p = 0.23$ [\onlinecite{Cooper2009}].
The critical doping below which an upturn appear in the resistivity of LSCO is therefore $p = 0.18$.

Boebinger and co-workers also performed measurements of the Hall coefficient \RH, again in fields up to 60~T, 
in both \BiONE~(Bi-2201) [\onlinecite{Balakirev2003}] and LSCO [\onlinecite{Balakirev2009}].
These revealed a small anomaly at low $T$, in the form of a peak in the (positive) Hall number \nH~$\sim 1/$\RH, located at $p \simeq 0.17$. 
An explanation for this anomaly has yet to be found.

%%%%%%

Starting in 2007, high-field Hall measurements in \YBCO~(YBCO), also up to 60~T, revealed that \RH~is deeply negative at $T \to 0$ 
in the doping range $0.08 < p < 0.16$ \cite{LeBoeuf2007, LeBoeuf2011}. 
Quantum oscillations observed in that same range [\onlinecite{Doiron-Leyraud2007,Sebastian2010,Ramshaw2015}] have been interpreted in terms of a small electron pocket in the Fermi surface, 
attributed to a reconstruction caused by some density-wave order \cite{Taillefer2009, Laliberte2011}.
Subsequent studies showed that this Fermi-surface reconstruction (FSR) is caused by charge-density-wave (CDW) modulations, 
detected by nuclear magnetic resonance (NMR) [\onlinecite{Wu2011,Wu2013,Wu2015}] and x-ray diffraction (XRD) [\onlinecite{Ghiringhelli2012,Chang2012a,Achkar2012}]
in the same doping range \cite{Hucker2014,Blanco-Canosa2014}.

%%%%%%

Recently, Hall measurements in YBCO were extended to higher doping by using fields up to 88~T [\onlinecite{Badoux2016}].
Two findings were made. 
First, the FSR ends at $p = 0.16 \pm 0.005$, as does the CDW phase (in zero field) \cite{Hucker2014,Blanco-Canosa2014}.
This means that the critical doping for CDW order, \pcdw~$= 0.16 \pm 0.005$, is distinctly lower than the pseudogap critical point, 
which in YBCO is located at \pstar~$= 0.19 \pm 0.01$ [\onlinecite{Tallon2001}].
A similar separation of normal-state critical points was also found in LSCO from high-field Seebeck measurements [\onlinecite{Badoux2016a}], 
with \pcdw~$= 0.15 \pm 0.005$ and \pstar~$=  0.18$ [\onlinecite{Laliberte2016}].
The implication is that the pseudogap phase 
is distinct from the CDW phase.
The pseudogap is not a high-temperature precursor of the low-temperature charge order, for example.
However, CDW order may well be a secondary instability of the pseudogap phase,
once the latter has set in [\onlinecite{Cyr-Choiniere2017}].

%%%%%%%%%%%%%%%%%%%%%%%%%%%%%%%%%  Figure 1   %%%%%%%%%%%%%%%%%%%%%%%%%%%%%%%%

\begin{figure}[t]
\centering
\includegraphics[width=8.5cm]{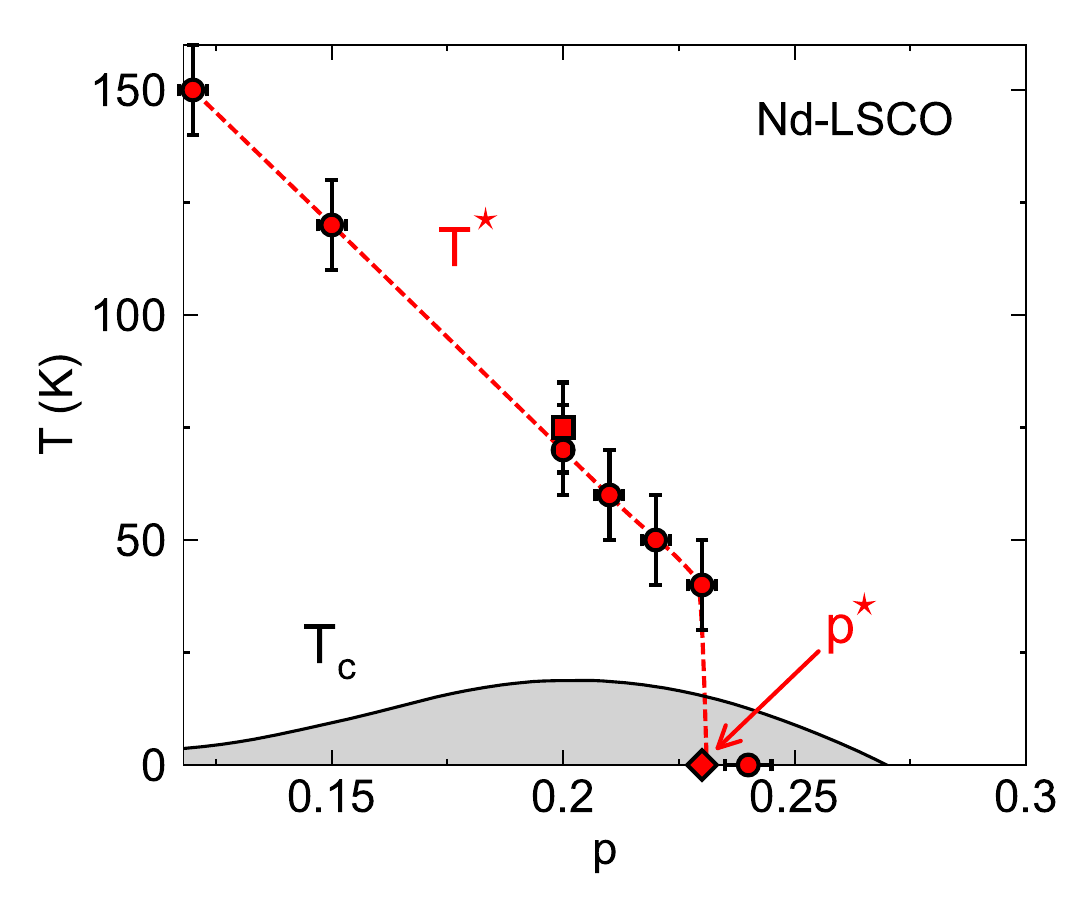}
\caption{
Temperature-doping phase diagram of Nd-LSCO,
showing the superconducting phase (grey) below \Tc~(black line) [\onlinecite{Daou2009}].
 The circles mark the onset of the upturn in the resistivity $\rho(T)$,
 %as reported in Ref.~\onlinecite{Daou2009} (white circles) and
 as observed in our data for $p = 0.20$, 0.21, 0.22, 0.23, and 0.24 (Figs.~2 and 4),
 and in the data of Ref.~\onlinecite{Ichikawa2000} for $p = 0.12$ and $p = 0.15$.
 The dashed red line is a guide to the eye ending on the $T=0$ axis at $p =$~\pstar~$=0.23$,
 the critical doping for the onset of the resistivity upturn (inset of Fig.~4). 
 The red square at $p = 0.20$ is the onset temperature for the opening of the pseudogap in Nd-LSCO, 
 as measured by ARPES [\onlinecite{Matt2015}].
 At $p=0.24$, the same ARPES study detects no pseudogap, down to \Tc[\onlinecite{Matt2015}].
 We can therefore identify the red dashed line as the pseudogap temperature \Tstar,
 and \pstar~as the $T=0$ critical point of the pseudogap phase (red diamond).
 %
% The temperature \Tsdw~(blue triangles) marks the onset of 
%spin-density-wave (SDW) modulations detected by neutron diffraction.\cite{Tranquada1997}
%The solid blue line is a smooth fit through the data points and the dashed blue line is
%a linear extrapolation to \Tsdw~$=0$.}
}
\label{Diagram}
\end{figure}

%%%%%%%%%%%%%%%%%%%%%%%%%%%%%%%%%%%%%%%%%%%%%%%%%%%%%%%%%%%%%%%%%%%%%%%

The second finding in YBCO is a dramatic drop in \nH~as doping is decreased below \pstar[\onlinecite{Badoux2016}].
This drop was attributed to a decrease in carrier density $n$, from $n = 1 + p$ at $p >$~\pstar~to $n = p$ at $p <$~\pstar. 
Based on this insight, it was recently demonstrated that the upturn in the resistivity of LSCO 
can be accounted for quantitatively, thereby resolving the 20-year-old puzzle [\onlinecite{Laliberte2016}].
The ``metal-to-insulator crossover" is in fact the consequence of a $T=0$ metal-to-metal transition into the pseudogap phase at \pstar,
whose ground state is a metal with $n = p$ holelike carriers.

In this paper, 
we study a third cuprate, \Nd~(Nd-LSCO), known to exhibit an upturn in both
$\rho(T)$~and \RH$(T)$ [\onlinecite{Daou2009}].
As we shall see,
this is really what proves that the upturns are due to a loss of carrier density.
%}
%
An important advantage of Nd-LSCO is that the opening of the pseudogap 
measured spectroscopically [by angle-resolved photoemission spectroscopy (ARPES)] [\onlinecite{Matt2015}] coincides with the start of 
the upturn in $\rho(T)$ [\onlinecite{Daou2009}], 
as a function of doping and temperature,
thereby linking the transport anomalies 
directly to the pseudogap phase (Fig.~1).
We report a detailed investigation of the transition across the pseudogap critical point of Nd-LSCO,
\pstar~$= 0.23$ [\onlinecite{Cyr-Choiniere2010}],
based on high-field measurements of $\rho$ and \RH~at $p = 0.20$, 0.21, 0.22, 0.23 and 0.24.
We show that the upturns in both coefficients are quantitatively consistent with a carrier density $n$
that drops from $1 + p$ to $p$ across \pstar. 
We find that the transition proceeds via an intermediate regime whose width in doping is comparable to that observed in YBCO (Ref.~\onlinecite{Badoux2016})
and LSCO [\onlinecite{Laliberte2016}].
Comparing to calculations [\onlinecite{Storey2016}] strongly suggests that the Fermi-surface transformation in these three cuprates
is caused by the sudden onset -- at a $T = 0$ critical point -- of a new 
Brillouin zone (or umklapp surface) akin to that produced by
the onset of an antiferromagnetic phase with wave vector $Q = (\pi, \pi)$.
In such a model, the width in \nH~vs $p$ is due to an intermediate regime in which the Fermi surface contains both holelike and electronlike carriers.
This offers a possible explanation for the puzzling Hall anomaly seen in LSCO [\onlinecite{Balakirev2009}].
%

%%%%%%%%%%%%%%%%%%%%%%%%%%%%%%%%%  Figure 2   %%%%%%%%%%%%%%%%%%%%%%%%%%%%%%%%

\begin{figure}[t]
\centering
\includegraphics[width=8.5cm]{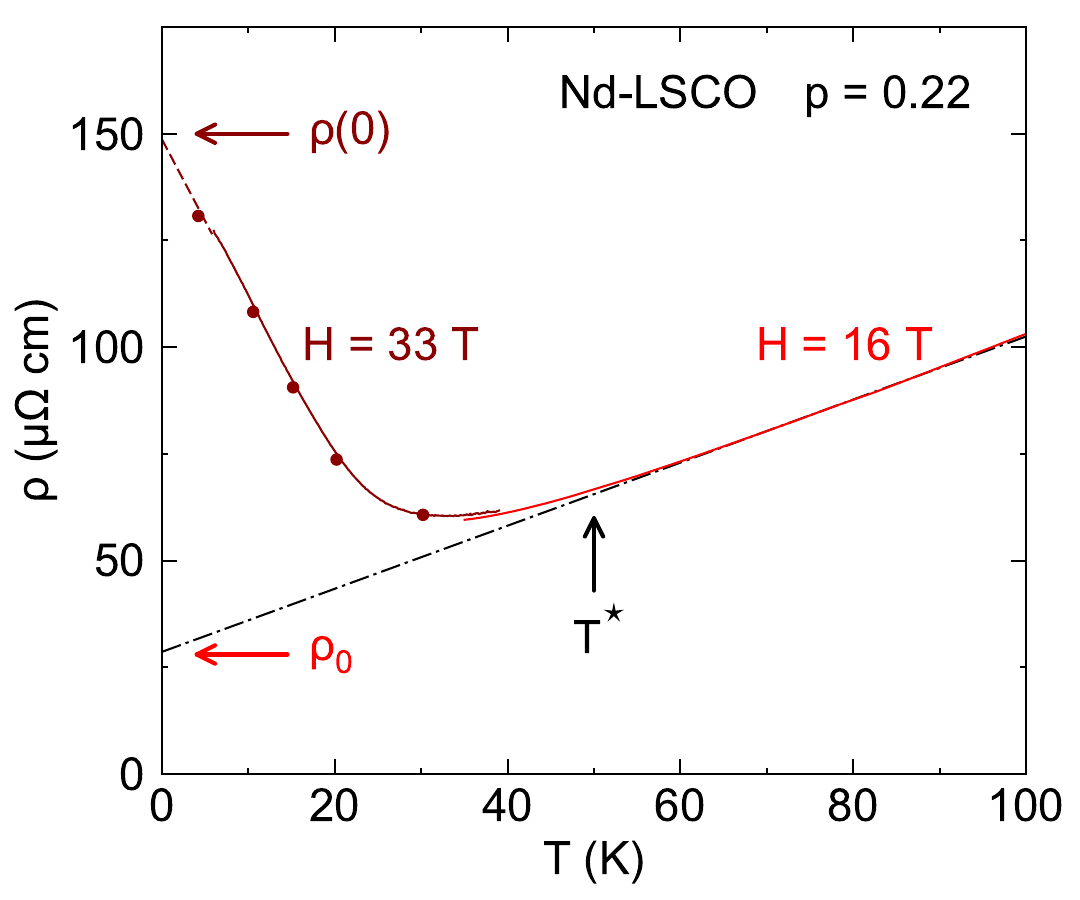}
\caption{
Electrical resistivity of Nd-LSCO at $p = 0.22$, as a function of temperature for two values of the magnetic field:
$H = 16$~T (red) and $H = 33$~T (burgundy).
The dots are obtained from the isotherms in Fig.~3b, taken at $H = 33$~T.
The straight dash-dotted line is a linear fit to the 16~T curve 
above 70~K,
which extrapolates to $\rho_0 = 29~\mu \Omega$~cm 
at $T$=0.
The measured curve is seen to deviate from this linear dependence below \Tstar~$\simeq 50$~K (arrow).
\Tstar~is the pseudogap temperature, plotted on the doping phase diagram in Fig.~1.
The burgundy dashed line is a linear extension of the 33~T curve below 7 K,
which yields $\rho(0) = 148~\mu \Omega$~cm at $T = 0$.
Correcting for the positive magneto-resistance [Fig.~9(b)] 
gives $\rho(0) = 136~\mu \Omega$~cm. 
}
\label{R-temperature}
\end{figure}

%%%%%%%%%%%%%%%%%%%%%%%%%%%%%%%%%%%%%%%%%%%%%%%%%%%%%%%%%%%%%%%%%%%%%%%

%%%%%%%%%%%%%%%%%%%%%%%%%%%%%%%%%  Figure 3   %%%%%%%%%%%%%%%%%%%%%%%%%%%%%%%%

\begin{figure*}[t]
\centering
\includegraphics[width=14cm]{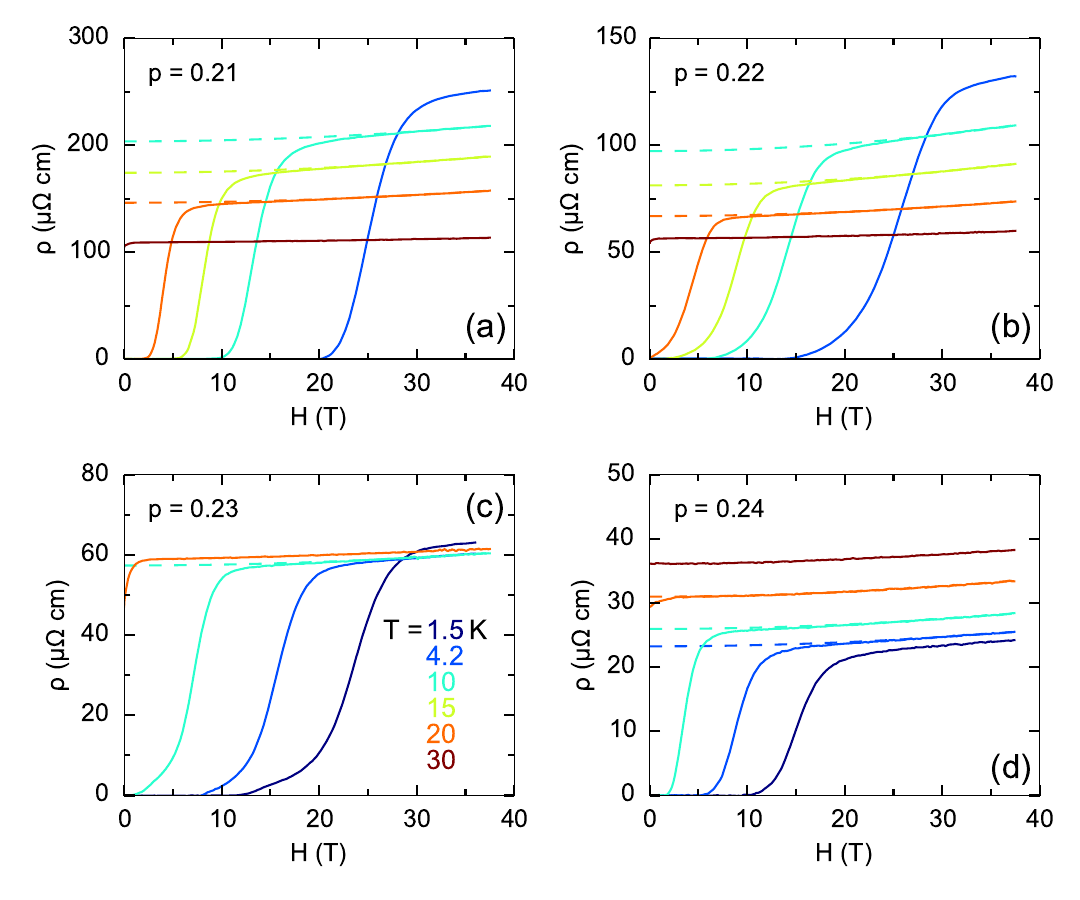}
\caption{
 Isotherms of the resistivity $\rho$ of Nd-LSCO as a function of magnetic field $H$,
for 4 dopings as indicated,
 at various temperatures as indicated.
 The dashed lines are $H^2$ fits to the normal-state data above the superconducting transition,
 which extrapolate to a value $\rho(H \to 0)$ at $H=0$.
}
\label{R-isotherms}
\end{figure*}

%%%%%%%%%%%%%%%%%%%%%%%%%%%%%%%%%%%%%%%%%%%%%%%%%%%%%%%%%%%%%%%%%%%%%%%

%%%%%%%%%%%%%%%%%%      METHODS

\section{Methods}

%Five samples ...
%Single crystal growth... 
%Contacts...
%\Tc~values...
%Doping values...
%Current direction...
%Field direction...
%\Hc~values...
%Measurements in Sherbrooke...
%Measurements in Nijmegen...

Large single crystals of Nd-LSCO were grown by a traveling float-zone technique in an image furnace, 
with nominal Sr concentrations $x = 0.20$, 0.21, 0.22, 0.23 and 0.25.
Samples were cut into small rectangular platelets of typical dimensions of 1~mm in length 
and 0.5~mm in width (in the basal plane of the tetragonal structure), with a thickness of 0.2~mm
along the $c$~axis.
The hole concentration $p$ of each sample is taken to be $p = x$, except for the $x = 0.25$ sample, whose doping
is $p = 0.24 \pm 0.005$ (see the Appendix).
Each sample is labeled by its $p$ value.
%\textcolor{red}{The hole concentration $p$ of each sample has been determined via their $T_c$ values, and their Hall coefficient values at $T = 80$ K.
%Within error bars, we found $p$ values close to the Strontium content $x$ except for the $x=0.25$ sample which yield $p = 0.236 \pm 0.002$.
%To soften the reasoning, we label the samples $p$ = 0.20, 0.21, 0.22, 0.23 and 0.24.
%One can refer to Annex B. for the determination of $p$, its exact values and error bars.}

Six contacts were made on each sample with H20E silver epoxy diffused by annealing at high temperature in oxygen
(two contacts for the current, two for the longitudinal resistivity and two for the transverse Hall signal).
Measurements were performed using a standard four-point technique with the current 
applied along the length of the sample (in the CuO$_2$ plane). 
The magnetic field was applied parallel to the $c$ axis (normal to the CuO$_2$ plane). 
All samples were measured in Sherbrooke at a fixed field of $H = 0$ and $H = 16$~T.
In Nijmegen, two types of measurements were carried out:
field sweeps up to 37.5~T at a typical speed of 4 T/min, at various fixed temperatures;
temperature sweeps at a fixed field of $H = 33$~T.

%%%%%%%%%%%%%%%%%%       RESULTS

\section{Results}

The aim of our study was to investigate in detail the onset of the pseudogap phase, 
as the material is taken across \pstar~(Fig.~1),
in the absence of superconductivity,
by measuring the in-plane transport coefficients $\rho$ and \RH~in magnetic fields large enough to access the normal state
at low temperature, for closely spaced dopings from $p = 0.20$ to $p = 0.24$.

\subsection{Resistivity}

At $p = 0.24$, $\rho(T)$ is known to be perfectly linear below 80~K [\onlinecite{Daou2009}].
Below \pstar, an upturn appears at low temperature.
In Fig.~2, the electrical resistivity of our Nd-LSCO sample with $p = 0.22$ is plotted as a function of temperature.
Above 50~K, $\rho(T)$ is linear in temperature.
Upon cooling below $T \simeq 50$~K, $\rho(T)$ shows a clear upturn at low $T$.
A temperature sweep at $H = 33$~T allows us to track that upturn down to $T \simeq 5$~K.
%

%%%%%%%%%%%%%%%%%%%%%%%%%%%%%%%%%  Figure 4   %%%%%%%%%%%%%%%%%%%%%%%%%%%%%%%%

\begin{figure}[t]
\centering
\includegraphics[width=8.5cm]{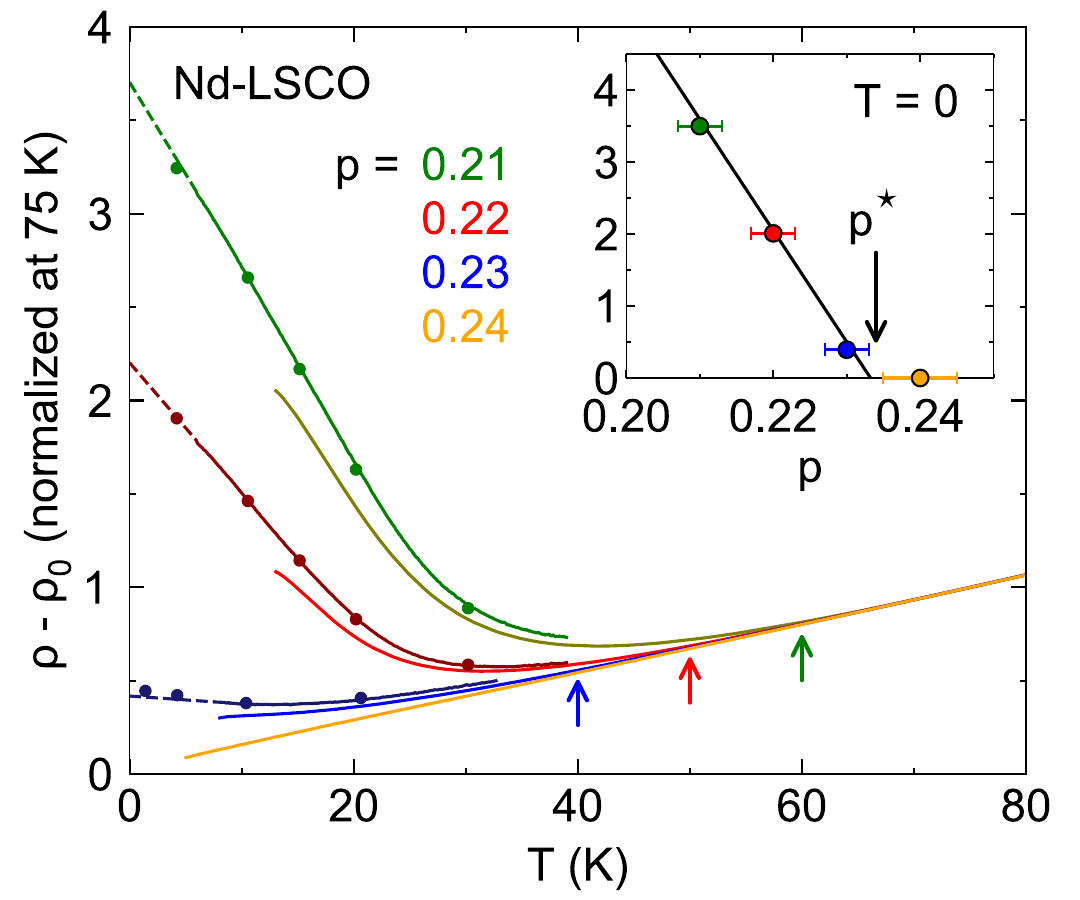}
\caption{
Doping evolution of the upturn in the normal-state resistivity of Nd-LSCO.
The temperature-dependent part of the resistivity, $\rho(T) - \rho_0$, is normalized to its value 
at $T = 75$~K for each sample.
The values of $\rho_0$ are obtained from linear fits as shown in Fig.~2 for $p = 0.22$.
Pairs of continuous curves are shown:
the lower curve (pale) is at $H = 16$~T,
the higher curve (dark) is at $H = 33$~T.
The dots are obtained from isotherms in Fig.~3, taken at $H = 36$~T.
The dashed lines are a 
linear extrapolation to $T = 0$ of the 33~T curves, yielding the (normalized) value
of $\rho(0) - \rho_0$ at each doping.
The color-coded arrows mark the onset of the upward deviation in $\rho(T)$ from its linear $T$ dependence
at high temperature, for $p = 0.21$ (green), $p = 0.22$ (red), and $p = 0.23$ (blue).
These onset temperatures \Tstar~are plotted in the $T-p$ phase diagram of Fig.~1. 
Inset:
Doping dependence of the normalized $\rho(0) - \rho_0$.
%
%The point at $p = 0.20$ (purple) is obtained from data in Ref.~\onlinecite{Daou2009}.
%
The line is a linear fit through the data points at $p = 0.21$, 0.22 and 0.23.
Its extrapolation to zero is one way to estimate the critical doping \pstar~(arrow).
}
\label{R-upturns}
\end{figure}

%%%%%%%%%%%%%%%%%%%%%%%%%%%%%%%%%%%%%%%%%%%%%%%%%%%%%%%%%%%%%%%%%%%%%%%

%%%%%%%%%%%%%%%%%%%%%%%%%%%%%%%%%  Figure 5   %%%%%%%%%%%%%%%%%%%%%%%%%%%%%%%%

\begin{figure}[t]
\centering
\includegraphics[width=8.4cm]{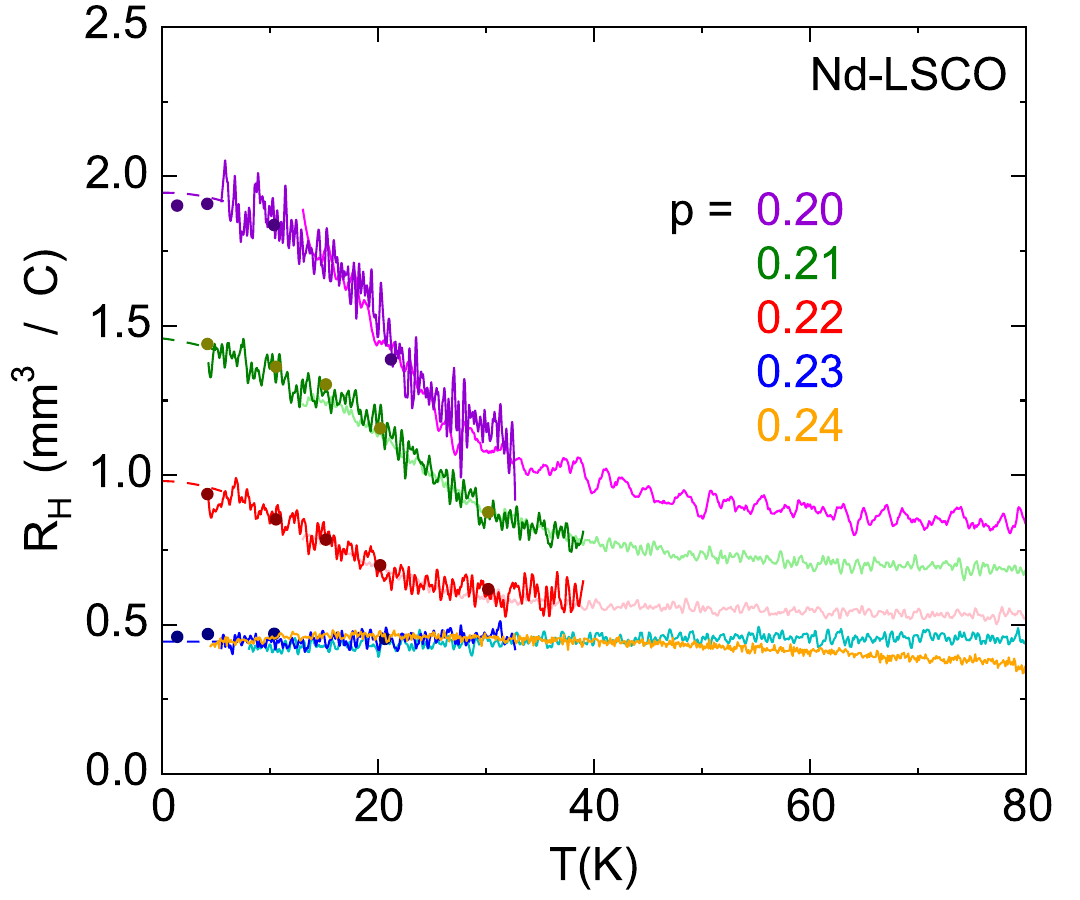}
\caption{
Temperature dependence of the Hall coefficient in Nd-LSCO, at five dopings as indicated.
The pale-colored curves were obtained at $H = 16$~T, the dark-colored ones (below 40~K) at $H = 33$~T.
The dots are obtained from the isotherms of Fig.~6, taken at $H = 36$~T. 
The dashed lines smoothly extrapolate the data to $T=0$, to obtain the value of \RH~at $T \to 0$,
\RH(0), at each doping.
%
%Note that the $p = 0.23$ curve lies above the $p=0.24$ curve at 80~K, but then merges with it at low temperature. 
}
\label{H-isotherms}
\end{figure}

%%%%%%%%%%%%%%%%%%%%%%%%%%%%%%%%%%%%%%%%%%%%%%%%%%%%%%%%%%%%%%%%%%%%%%%

In Fig.~3, we report several low-$T$ isotherms of $\rho$ vs $H$ measured up to $H = 37.5$~T
in our samples of Nd-LSCO with $p = 0.21$, 0.22, 0.23 and 0.24.
We see that by 33~T the normal state is reached at all temperatures down to at least 4~K.
The temperature dependence can be obtained by taking a cut at fixed field.
Doing this for $p = 0.22$ at $H = 33$~T yields the dots plotted in Fig.~2, in good agreement
with the continuous 33~T curve. 

It is useful to characterize the resistivity of Nd-LSCO in two ways.
First, by fitting the linear regime above 70~K to $\rho(T) = \rho_0 + A T$ (dash-dotted line in Fig.~2),
we obtain the extrapolated residual resistivity $\rho_0$.
Second, by extending the high-field low-temperature 
data down to $T=0$ (dashed line in Fig.~2), we obtain
the actual normal-state resistivity in the $T = 0$ limit,
$\rho(0)$.
At $p = 0.22$, for example, we obtain
$\rho_0 = 29 \pm 2~\mu \Omega$~cm and
$\rho(0) = 148 \pm 3~\mu \Omega$~cm (Fig.~2).

%%%%%%%%%%%%%%%%%%%%%%%%%%%%%%%%%  Figure 6   %%%%%%%%%%%%%%%%%%%%%%%%%%%%%%%%

\begin{figure*}[t]
\centering
\includegraphics[width=14cm]{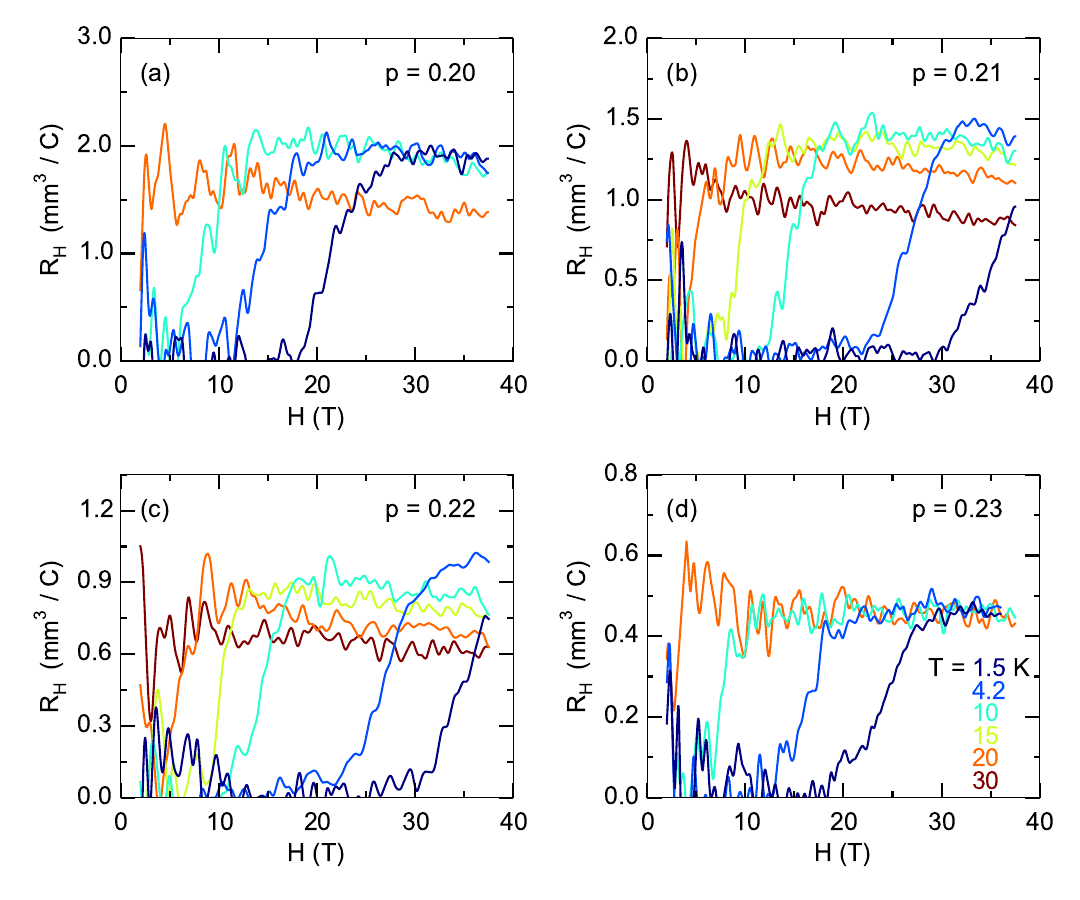}
\caption{
Isotherms of the Hall coefficient in Nd-LSCO, 
as a function of magnetic field $H$, for four dopings as indicated,
at various temperatures as indicated.
}
\label{H-isotherms}
\end{figure*}

%%%%%%%%%%%%%%%%%%%%%%%%%%%%%%%%%%%%%%%%%%%%%%%%%%%%%%%%%%%%%%%%%%%%%%%

There is a positive magnetoresistance (MR) in all samples, which grows as $H^2$ (Fig.~9).
By extrapolating to $H = 0$ a quadratic fit to the high-field data in the normal state 
(dashed lines in Fig.~3),
we obtain $\rho(H\to0)$,
and define the relative MR
as $\Delta \rho / \rho(H\to0) \equiv \rho(H) / \rho(H\to0) - 1$.
In Sec.~III.~D, we relate this MR to the mobility, studied across \pstar.
Using MR data as in Fig.~9(b) for $p=0.22$, we can remove the MR from the value of $\rho(0)$ obtained in high fields (\eg~33~T).
For $p=0.22$, this yields 
$\rho(0) = 136 \pm 5~\mu \Omega$~cm.
We use the dimensionless ratio of this MR-free value of $\rho(0)$ to $\rho_0$ to
quantify the change in resistivity at $T=0$ caused by the pseudogap (see Sec.~III.~C).

%\textcolor{red}{Annex A shows how magnetoresistance is taken into account to correct the $\rho (T \rightarrow 0)$ for $p=0.21$, 0.22 and 0.23.
%As we lack field sweeps for $p=0.20$, we estimated a MR at $T=0$ from the three dopings shown figure \ref{MR} which yield a MR varying between 4 and 7\% at $H = 33 $ T.
%This procedure is not necessary for $p = 0.24$ as only a small 16 T field is needed to suppress superconductivity down to 5 K ; at this field the MR is negligible (of the order of the percent).
%}

In Fig.~4, we show how the upturn in $\rho(T)$ evolves with doping.
For a close comparison of data from four different samples,
we normalize the four curves so that they are all equal above 60~K.
Specifically, we subtract $\rho_0$ and then normalize $\rho(T) - \rho_0$ to unity at $T = 75$~K.
%
%Data obtained from three separate measurements are plotted:
%a temperature sweep at $H = 16$~T (pale lines), a temperature sweep at $H = 33$~T (dark lines), 
%and data points at $H = 36$~T (dots, from isotherms in Fig.~3).
%
At $p = 0.24$, the data show that $\rho(T)$ is perfectly linear below 80~K, 
as observed before [\onlinecite{Daou2009}].
As soon as the doping is reduced below \pstar, an upturn in $\rho(T)$ develops at low $T$.
The upturn grows rapidly as $p$ is further reduced.
The value of $\rho(T) - \rho_0$ at $T \to 0$ is plotted in the inset of Fig.~4.
We see that it grows linearly from \pstar~down, 
in good agreement with 
the $c$-axis resistivity [\onlinecite{Cyr-Choiniere2010}],
thereby confirming, on a different set of samples, the location of the critical doping in Nd-LSCO, 
at \pstar~$= 0.23 \pm 0.01$.

\subsection{Hall coefficient}

The Hall coefficient \RH~was measured 
simultaneously
on the same samples as the resistivity.
The data for five samples with dopings $p = 0.20$, 0.21, 0.22, 0.23 and 0.24 are displayed
in Fig.~5 as a function of temperature.
Curves at $H = 16$~T reveal the essential features, confirmed and extended to lower $T$ by the 33-T curves.
The data at $p = 0.20$ and $p = 0.24$ are in excellent agreement with the previous study [\onlinecite{Daou2009}].
Isotherms up to $H = 37.5$~T are displayed in Fig.~6.
These show that the normal state is reached at $H = 33$~T for all temperatures down to $T = 4.2$~K
for $p = 0.21$ and 0.22, and down to $T = 1.5$~K for $p = 0.20$ and 0.23.
Cuts at $H = 33$~T agree very well with the temperature sweeps of Fig.~5.

At $T = 80$~K, \RH~increases monotonically with decreasing $p$, as it does in all
hole-doped cuprates at $T >$~\Tstar[\onlinecite{Ando2004a,Segawa2004}].
At $p = 0.24$, as observed before [\onlinecite{Daou2009}],
\RH$(T)$ is flat below $\sim 50$~K, and \RH$(0) \simeq V / e (1+p)$, where $V$ is the unit-cell volume and $e$ is the electron charge,
in agreement with the value expected for
a single large holelike Fermi surface containing $1 + p$ holes per Cu atom.

At $p = 0.20$, 0.21 and 0.22, there is a clear upturn in \RH$(T)$ at low $T$, starting roughly below 
the temperature where $\rho(T)$ has its minimum (Fig.~4).
In other words, the upturn in $\rho(T)$ also shows up in \RH$(T)$.
However, this is not true at $p = 0.23$, where \RH$(T)$~shows no upturn at low $T$ (Fig.~5).
%
%Furthermore, while \RH~grows monotonically with decreasing $p$ at $T = 80$~K (Fig.~5),
%so that its value at $p = 0.23$ lies between that of $p = 0.24$ and $p = 0.22$,
%this is no longer true at $T \to 0$, where \RH$(p=0.23) \simeq$~\RH$(p=0.24)$.
%
%In other words, something is reducing \RH$(T)$ at $p = 0.23$ 
%upon cooling, relative to \RH$(T)$ at $p = 0.24$.
%
%In Sec.~IV.~C, we argue that this reduction is caused by electronlike carriers in the Fermi surface
%of Nd-LSCO immediately below \pstar.

%%%%%%%%%%%%%%%%%%%%%%%%%%%%%%%%%  Figure 7   %%%%%%%%%%%%%%%%%%%%%%%%%%%%%%%%

\begin{figure}[t]
\centering
\includegraphics[width=9.0cm]{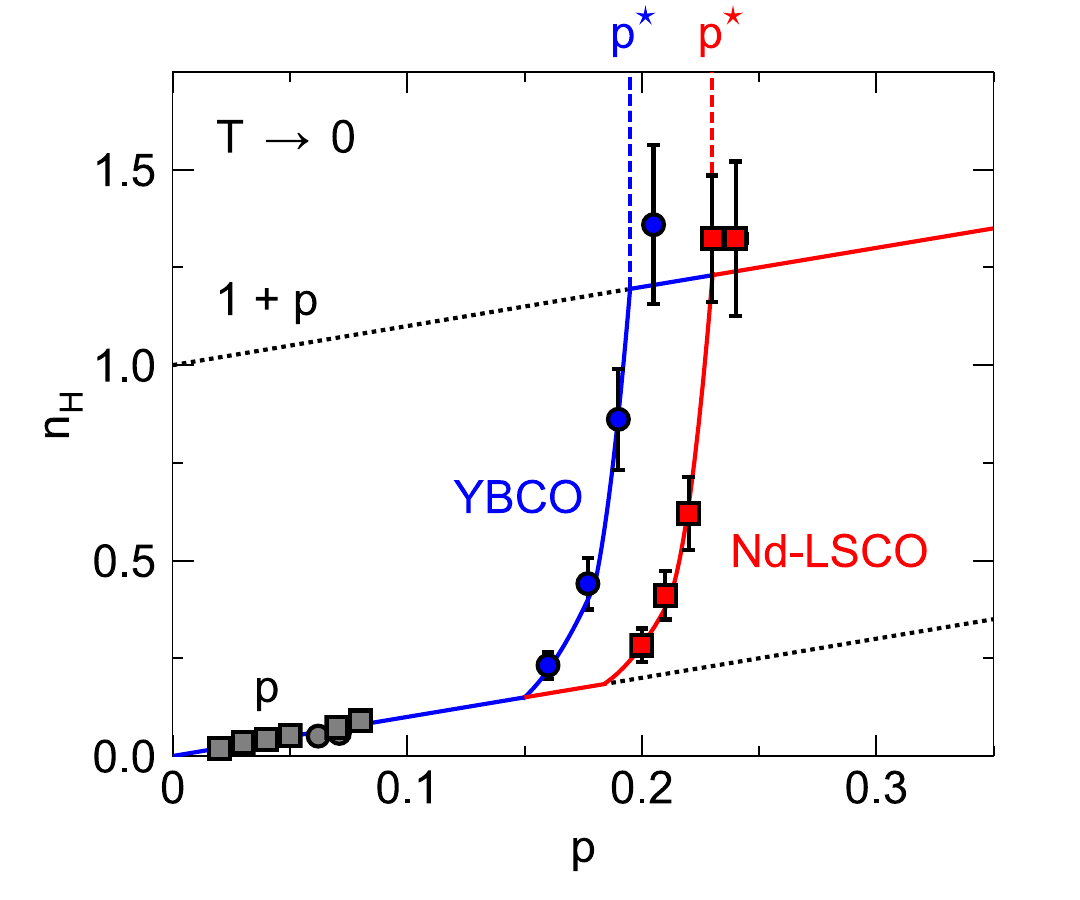}
\caption{
Hall number \nH~at $T \to 0$ as a function of doping, for Nd-LSCO (red squares) and YBCO (blue circles) [\onlinecite{Badoux2016}].
Below $p = 0.1$, the grey squares are for LSCO [\onlinecite{Ando2004}],
and the grey circles for YBCO [\onlinecite{Segawa2004}].
The vertical dashed lines mark the location of the pseudogap critical point, 
at \pstar~$= 0.23 \pm 0.01$ in Nd-LSCO (red)
and \pstar~$= 0.195 \pm 0.01$ in YBCO (blue) [\onlinecite{Badoux2016}].
The solid blue and red lines are a guide to the eye.
%
%The solid blue line is a guide to the eye. 
The two dotted lines mark \nH~$=1+p$ and \nH~$=p$, as indicated.
}
\label{YBCO-Nd-LSCO}
\end{figure}

%%%%%%%%%%%%%%%%%%%%%%%%%%%%%%%%%%%%%%%%%%%%%%%%%%%%%%%%%%%%%%%%%%%%%%%

\subsection{Carrier density}

In Fig.~7, we plot the Hall number \nH~$= V/(e$\RH) at $T \to 0$ as a function of doping,
obtained using \RH(0), the value of \RH~extrapolated to $T=0$ in Fig.~5. 
We see that at $T=0$ the onset of the pseudogap at \pstar~causes a drop from 
\nH~$\simeq 1+p$ at $p >$~\pstar~to \nH~$\simeq p$ at $p <$~\pstar.
%
%\sout{Note, however, that this drop in \nH~does not start immediately below \pstar~$=0.235$:
%it is not detected at $p = 0.23$, but only at $p = 0.22$.}

At $p=0.24$, it is certainly reasonable to interpret \nH~as a carrier density (with units of holes/Cu atom),
since the data yield \nH(0)~$= 1.3 \pm 0.1$
and the Luttinger rule requires the carrier density to be $n = 1+p = 1.24$ 
for a single large holelike Fermi surface.
By itself, the drop in \nH~below \pstar~does not necessarily imply a drop in carrier density,
for it could be due to a change in Fermi-surface curvature,
such as could occur at a nematic quantum critical point [\onlinecite{Nie2014}].
However, the fact that $\rho(T)$ shows an increase at low $T$, does imply 
a loss of carriers.
A drop of carrier density from $1+p$ to $p$ will cause the resistivity
at $T=0$ to increase by a factor $(1+p)/p$,
if the mobility does not change
(we show in Sec.~III.~D that it changes very little).
It is remarkable that this factor is precisely what is observed in Nd-LSCO, 
as noted earlier for $p = 0.20$ [\onlinecite{Daou2009}],
in the sense that the resistivity at $T \to 0$, $\rho(0)$,
is larger than the residual resistivity the metal would have at that doping,
$\rho_0$,
 if the pseudogap did not cause an upturn.
 Indeed, at $p = 0.20$, $\rho(0)/\rho_0 = 5.8$ [\onlinecite{Daou2009}],
 while $(1+p)/p = 6$.

 Following Ref.~\onlinecite{Laliberte2016}], we define the carrier density \nrho~derived from $\rho(T)$,
 as \nrho~$\equiv (1+p) [\rho_0/\rho(0)]$.
 By construction, this gives \nrho~$= 1 + p$ at $p = 0.24$ since at that doping there is no upturn,
 and $\rho(0) = \rho_0$.
In Fig.~8, we plot \nrho~vs $p$ [using MR-corrected values of $\rho(0)$] and
 see that \nrho~$\simeq p$, 
 %within error bars, 
 at $p = 0.20$, 0.21 and 0.22.
%
%\sout{Note that correcting for the positive MR will increase \nrho~by at most 11~\% (see below),
%which is within the error bars of the data points in Fig.~8a.}
Note, that the drop in \nrho~starts earlier than the drop in \nH.
In Sec.~IV.~C, we mention a possible explanation for this difference.
%\textcolor{red}{All our $\rho (0)$ values have been corrected to compensate the MR as explained Annex A.} 

%%%%%%%%%%%%%%%%%%%%%%%%%%%%%%%%%  Figure 8   %%%%%%%%%%%%%%%%%%%%%%%%%%%%%%%%

\begin{figure}[t]
\centering
\includegraphics[width=9.4cm]{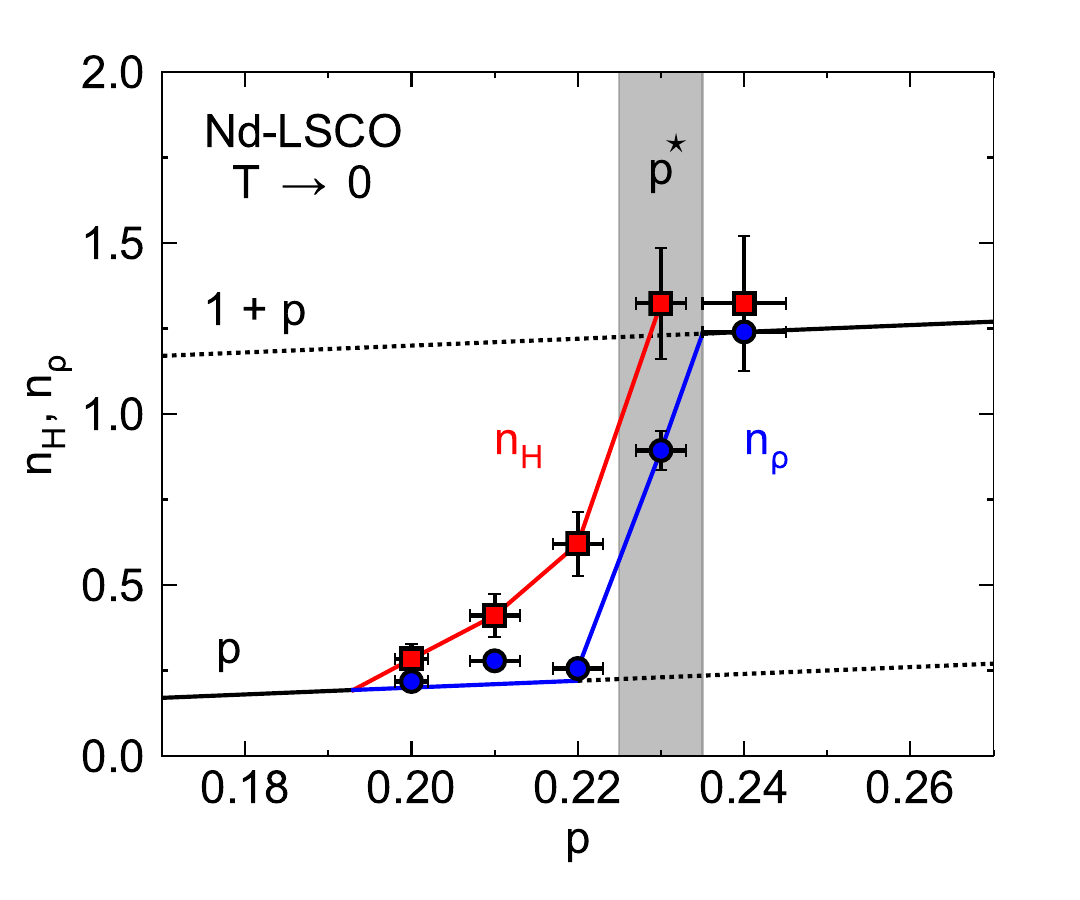}
\caption{
%
%(a)
Doping dependence of the carrier density in Nd-LSCO, estimated in two ways:
1) from the Hall effect, as \nH~$=V /e$\RH(0)~(red squares; Fig.~7); 
2) from the resistivity,
as \nrho~$\equiv (1+p) \rho_0 / \rho(0)$~(blue circles),
where $\rho(0)$ is corrected for the magnetoresistance (see text).
In both cases, the normal-state data in the $T=0$ limit are used.
The vertical grey band marks the location of the pseudogap critical point \pstar~$= 0.23$.
The upper dotted line marks $n =1+p$; the lower dotted line marks $n = p$. 
%
%
%(b)
%Ratio of \nH~over \nrho~(squares), obtained from data points in the upper panel.
%
%In both panels, 
The black, blue, and red solid lines are a guide to the eye.
}
\label{density}
\end{figure}

%%%%%%%%%%%%%%%%%%%%%%%%%%%%%%%%%%%%%%%%%%%%%%%%%%%%%%%%%%%%%%%%%%%%%%%

%%%%%%%%%%%%%%%%%%%%%%%%%%%%%%%%%  Figure 9   %%%%%%%%%%%%%%%%%%%%%%%%%%%%%%%%

\begin{figure}[t]
\centering
\includegraphics[width=8.5cm]{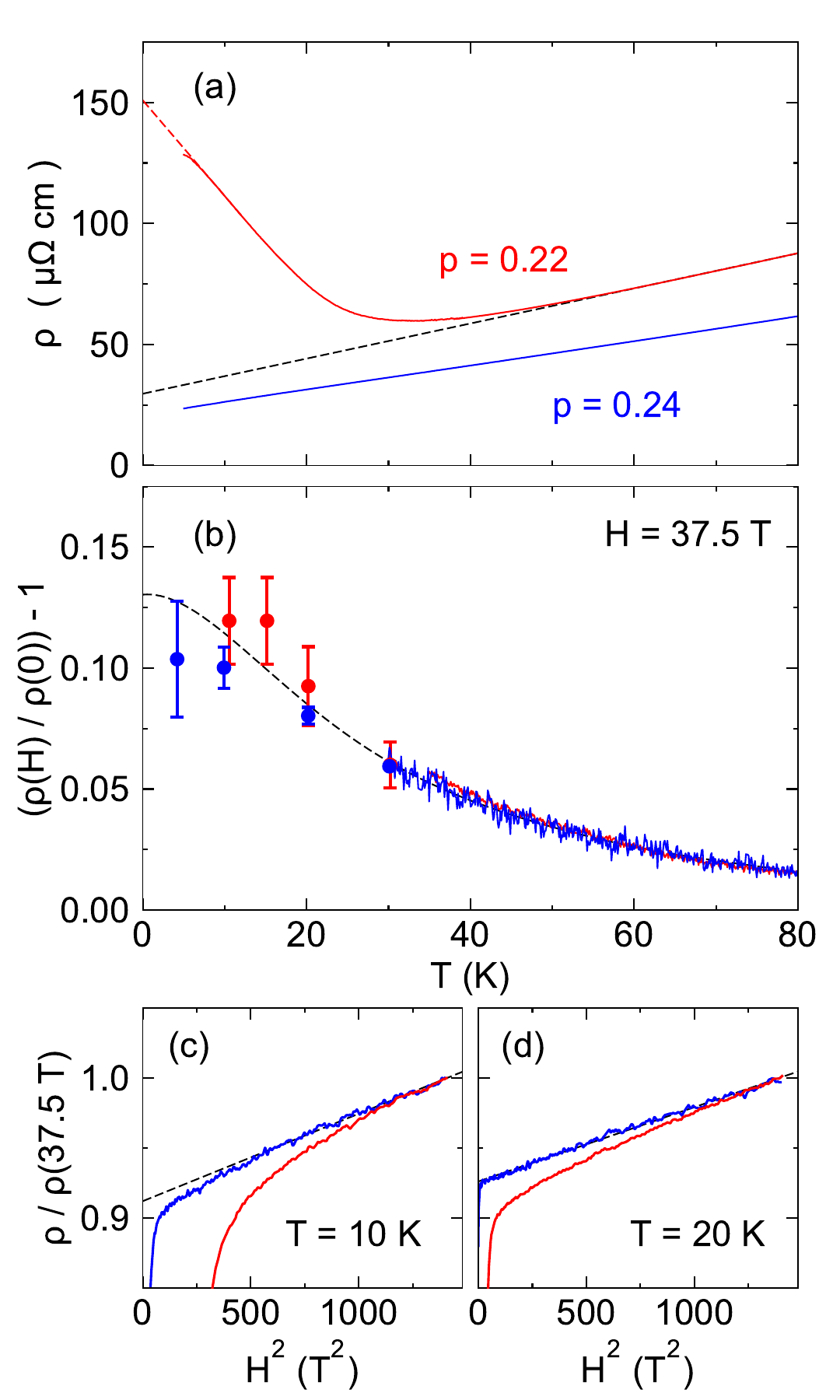}
\caption{
(a)
Temperature dependence of the field-induced normal-state resistivity of Nd-LSCO for a doping just 
above \pstar~($p=0.24$, blue) and one just below ($p=0.22$, red). 
A magnetic field of 33~T was applied to suppress superconductivity. 
The dashed line is a linear fit to the $p = 0.22$ data above $T = 70$~K. 
(b)
Relative magneto-resistance (MR) of the same two samples plotted as $[\rho(H) / \rho(H \to 0)]-1$ vs $T$, 
for $H = 37.5$~T (dots), where $\rho(H)$ and $\rho(H \to 0)$ 
are obtained from Fig.~3.
The red and blue curves are obtained from the MR at 16~T, scaled up to 37.5~T assuming that MR $\propto (\mu H)^2$. 
The dashed line is a guide to the eye.
(c)
Field dependence of the resistivity for $p = 0.22$ (red) and $p = 0.24$ (blue), at $T = 10$~K,
plotted as $\rho$ vs $H^2$, with $\rho$ normalized to its value at $H = 37.5$~T.
The black dashed line is a linear fit to the $p = 0.24$ data, at high field.
(d) Same as in (c), but for $T = 20$~K.
}
\label{mobility}
\end{figure}

%%%%%%%%%%%%%%%%%%%%%%%%%%%%%%%%%%%%%%%%%%%%%%%%%%%%%%%%%%%%%%%%%%%%%%%

%%%%%%%%%%%%%%%%%%%%%%%%%%%%%%%%%  Figure 10   %%%%%%%%%%%%%%%%%%%%%%%%%%%%%%%%

\begin{figure}[t]
\centering
\includegraphics[width=8.8cm]{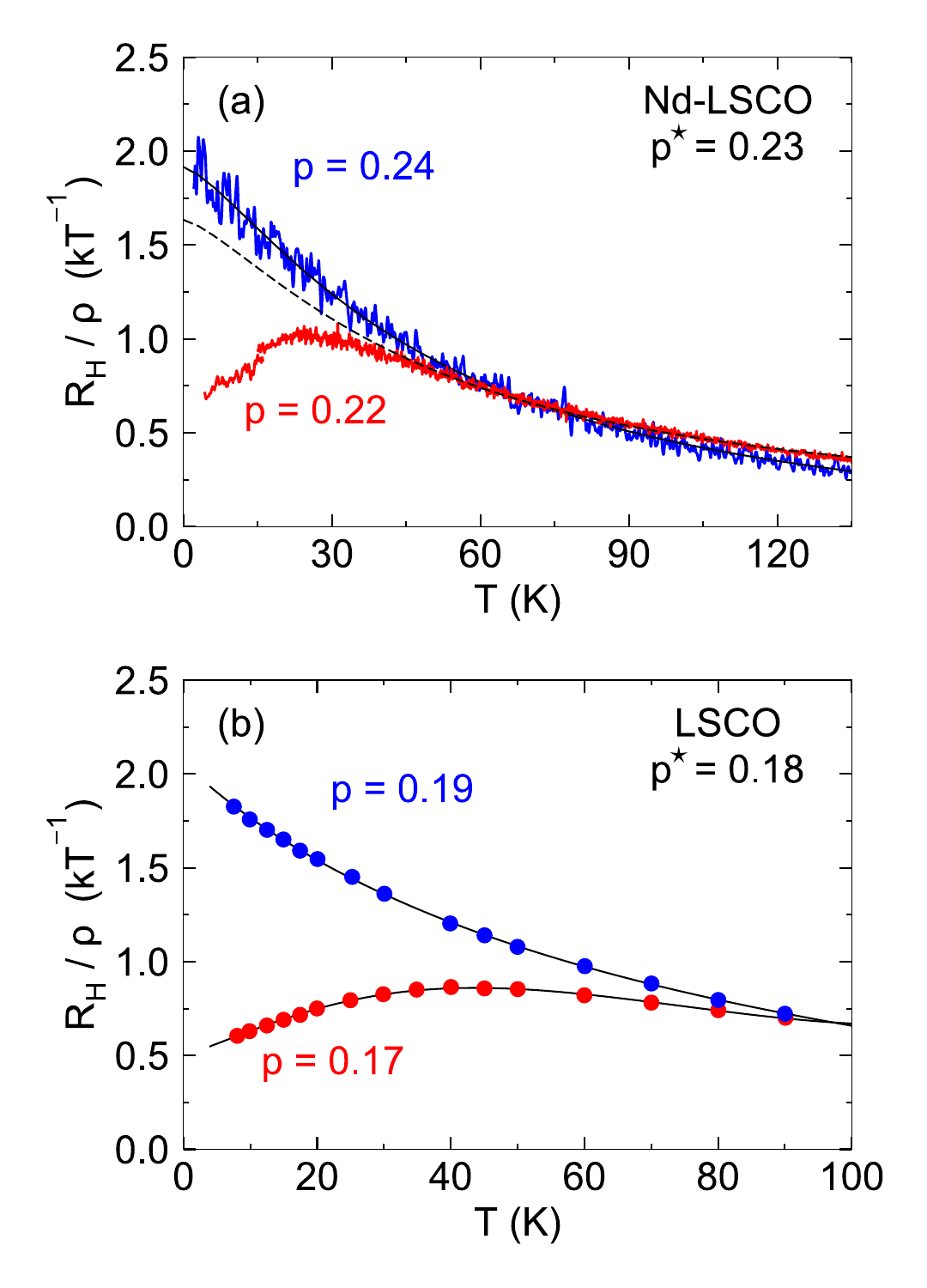}
\caption{
(a)
Ratio of \RH~(Fig.~5) over $\rho$~(Fig.~9a)
for our Nd-LSCO samples with $p = 0.22 <$~\pstar~(red) and $p = 0.24 >$~\pstar~(blue).
The solid line is a smooth fit through the $p = 0.24$ data.
The dashed line is a similar fit through the $p = 0.22$ data, 
above \Tstar~$\simeq 50$~K.
In a single-carrier situation, this ratio is proportional to the mobility.
We see that it is very similar for the two samples,
above 50~K, in agreement with the MR data (Fig.~9b).
For $p = 0.22$, the drop below $T \simeq 30$~K could come from an electronlike contribution
to the Hall signal (Sec.~IV.~C).
(b)
Same ratio for LSCO, at $p = 0.17 <$~\pstar~(red) and $p = 0.19 >$~\pstar~(blue),
calculated from published \RH($T$) data [\onlinecite{Balakirev2009}] and $\rho(T)$ data \cite{Boebinger1996,Cooper2009}.
Here the lines are a guide through the data points.
}
\label{Hall-angle}
\end{figure}

%%%%%%%%%%%%%%%%%%%%%%%%%%%%%%%%%%%%%%%%%%%%%%%%%%%%%%%%%%%%%%%%%%%%%%%

\subsection{Mobility}

It is instructive to investigate the impact of the pseudogap phase on the mobility $\mu$ of the charge carriers.
We estimate $\mu$ in two separate ways. First, by looking at the magneto-resistance,
which varies as MR~$\propto (\omega_{\rm c} \tau)^2 \propto (\mu H)^2$ in the weak-field limit,
where $\omega_{\rm c}$ is the cyclotron frequency and $\tau$ is the scattering time.
The MR in our data does vary as $H^2$ (Fig.~9).
In Fig.~9(b), we plot the relative MR as a function of temperature, evaluated at 37.5~T,
for two dopings, above and below \pstar.
At $p = 0.24 >$~\pstar, we see that the MR decreases monotonically with increasing temperature,
by a factor of $\sim 8$ between $T=0$ and $T = 80$~K [Fig.~9(b)].
This is roughly consistent with the three-fold increase in $\rho$ over that interval [Fig.~9(a)],
reflecting an increase in scattering rate (decrease in $\tau$) by a factor $\sim 3$.

Comparing the data at $p = 0.24$ with MR data at $p = 0.22 <$~\pstar~reveals a striking fact:
even though the resistivity now undergoes a huge upturn that causes a five-fold enhancement 
of its value at $T = 0$ [Fig.~9(a)], the MR is virtually identical to (perhaps even slightly larger than) 
what it was above \pstar~[Fig.~9(b)], 
\ie~it is essentially unaffected by the onset of the pseudogap.
This demonstrates two important facts. 
First, the upturn in the resistivity is not due to an increase in scattering rate. 
Second, the upturn reflects essentially the full drop in carrier density, without the usual compensating enhancement of the mobility across a transition, 
so that $\rho \sim 1 / (n e \mu) \sim 1 / n$.
The same observation, $\rho \sim 1 / n$, was made in LSCO [\onlinecite{Laliberte2016}].
This interesting property of the pseudogap phase provides a window on the nature
of disorder scattering in that phase.

The second way to estimate the mobility is through the Hall angle, 
controlled
by the ratio
\RH$/ \rho$, which is proportional to $\mu$ in a single-band (single-carrier) metal.
In Fig.~10(a), we plot the ratio \RH$/ \rho$ as a function of temperature for the same two Nd-LSCO samples.
At $p = 0.24 >$~\pstar, we see that \RH$/ \rho$ decreases monotonically with increasing temperature, 
by a factor 3 between $T=0$ and $T = 80$~K, consistent with the threefold increase in scattering rate
[given the flat \RH~(Fig.~5)].
Note that the value at $T=0$ is such that $\omega_{\rm c} \tau =$~\RH$H/ \rho = 0.075$
at $H = 37.5$~T. The fact that $\omega_{\rm c} \tau << 1$ shows that we are indeed 
in the weak-field limit, justifying the use of a $H^2$ fit for the MR.

The ratio \RH$/ \rho$ is very similar for the two samples,
above $T \simeq 30$~K, in agreement with the MR data [Fig.~9(b)].
Below 30~K, however, \RH$/\rho$ at $p = 0.22$ shows a pronounced drop [Fig.~10(a)],
not seen at all in the MR [Fig.~9(b)].
It is therefore not due to a change of mobility at low temperature.
As discussed in Sec.~IV.~C, 
this anomaly may reflect the onset of electronlike carriers
generated when the Fermi surface is transformed by the pseudogap phase.

%%%%%%%%%%%%%%%%%%       DISCUSSION

\section{Discussion}

\subsection{Pseudogap in ARPES and transport}

Daou \etal~attributed the upturn in $\rho(T)$ they measured in Nd-LSCO at $p = 0.20$
to the opening of the pseudogap without direct spectroscopic evidence [\onlinecite{Daou2009}].
Recently, Matt \etal~reported ARPES measurements on Nd-LSCO that confirm this interpretation [\onlinecite{Matt2015}].
They observe a partial anti-nodal gap at $p = 0.20$, in the normal state just above \Tc.
They track this pseudogap as a function of temperature and find that it closes at \Tstar$= 75 \pm 10$~K.
This is in excellent agreement with the value \Tstar$= 80 \pm 15$~K reported by Daou \etal~for the onset of the 
upturn in $\rho(T)$ at $p = 0.20$ [\onlinecite{Daou2009}],
and with our own data (Fig.~1).

At $p = 0.24$, Matt \etal~observe no gap at all, confirming that the pseudogap phase begins below $p = 0.24$ [\onlinecite{Matt2015}].
Again, this is perfectly consistent with transport data (Fig.~1).
Their ARPES study therefore establishes clearly that the upturn in $\rho(T)$ observed in Nd-LSCO
is a signature of the pseudogap phase.
The same link between the ARPES-detected pseudogap and the onset of resistivity upturn
has been made for LSCO \cite{Laliberte2016,Cyr-Choiniere2017}.

Note that the signature of \Tstar~in $\rho(T)$ can be different in different cuprates or samples.
While it is typically an upturn in samples of Nd-LSCO and LSCO \cite{Laliberte2016,Ando2004},
it is typically a downturn in YBCO (Ref.~\onlinecite{Ando2004}), for example.
We can understand this difference if the effect of the pseudogap is to cause not only
a loss of carrier density, which increases $\rho$,
but also a loss of inelastic scattering, which decreases $\rho$ [\onlinecite{Cyr-Choiniere2017}].
In clean samples, like typical YBCO samples, the latter effect dominates
and so $\rho(T)$ drops below \Tstar, whereas in typical samples of Nd-LSCO or LSCO,
which are more disordered, the magnitude of inelastic scattering is much smaller relative to the magnitude
of elastic disorder scattering, and so the loss of carrier density overwhelms any loss of inelastic scattering,
and $\rho(T)$ rises below \Tstar.
To see in YBCO a low-$T$ upturn in $\rho(T)$ one needs to introduce disorder,
as was done by Rullier-Albenque \etal~with electron irradiation [\onlinecite{Rullier-Albenque2008}].
The upturn they saw in the resistivity of YBCO at $p = 0.18$ is in quantitative agreement with the 
carrier density measured by the Hall effect [\onlinecite{Badoux2016}] (see Ref.~\onlinecite{Laliberte2016}).
In Bi-2201, both features are observed:
$\rho(T)$ shows a slight drop below \Tstar[\onlinecite{Ando2004}],
at a temperature consistent with the opening of the pseudogap seen in ARPES [\onlinecite{Kondo2011}],
and it also shows a pronounced upturn at $T \to 0$ [\onlinecite{Ono2000}].

In summary, the loss of carrier density detectable in transport properties is a generic signature of the 
critical point \pstar~at which the pseudogap opens in the normal state of cuprate superconductors at $T = 0$.

\subsection{CDW critical point}

Having established that \pstar~$= 0.23$ is the critical doping at which the pseudogap phase 
begins in the normal state of Nd-LSCO at $T \to 0$ (Fig.~1), we now need to identify the critical 
point \pcdw~where CDW order sets in.
Daou \etal~assumed that the stripe order seen in Nd-LSCO at low temperature also ended at \pstar,
\ie~that CDW and SDW modulations both ended at that point [\onlinecite{Daou2009,Daou2009a}].
Recent studies have found that
\pcdw~lies well below \pstar, so that CDW and pseudogap phases are distinct \cite{Badoux2016,Badoux2016a}.

Indeed, in YBCO, XRD studies find that CDW modulations vanish at \pcdw~$=0.16 \pm 0.005$ \cite{Hucker2014,Blanco-Canosa2014}.
High-field Hall data at $T \to 0$ reveal that \RH~$<0$ at $p = 0.15$ whereas \RH~$>0$ at $p = 0.16$,
showing that the CDW-induced FSR also ends at $p = 0.16$ [\onlinecite{Badoux2016}],
while various measurements of the pseudogap onset yield \pstar~$=0.19 \pm 0.01$ [\onlinecite{Tallon2001}],
and the drop in \nH~is seen at \pstar~$=0.195 \pm 0.005$ [\onlinecite{Badoux2016}].

In LSCO, high-field Seebeck data were used in a similar fashion to pin down the end point
of FSR, giving \pcdw~$= 0.15 \pm 0.005$ [\onlinecite{Badoux2016a}].
This is again consistent with the fact that XRD detects no CDW modulations in LSCO at $p = 0.15$ [\onlinecite{Croft2016}].
Given that it is observed in two rather different cuprate materials, the separation between 
\pstar~and \pcdw~is most likely a generic property of cuprates.
Note that the difference (\pstar~$-$~\pcdw)~$\simeq 0.03 - 0.04$ in both cases.

In Nd-LSCO, 
CDW order has been detected by XRD at $p = 0.15$ [\onlinecite{Niemoller1999}],
but there is no report of any CDW modulations at $p > 0.15$.
(Note that SDW modulations are seen with neutron diffraction at $p = 0.20$ [\onlinecite{Tranquada1997}],
but this does not necessarily imply the presence of CDW order.
Indeed, all cuprates show SDW modulations at $p < 0.08$, without any CDW modulations.)
Both Hall and Seebeck coefficients drop at low temperature (with $S < 0$) at $p = 0.15$ \cite{Noda1999,Hucker1998},
while both \RH$(T)$~and $S(T)$ grow monotonically as $T \to 0$ (and remain positive) at $p = 0.20$ \cite{Daou2009,Daou2009a}. 
Hence in Nd-LSCO, \pcdw~$< 0.20$, separated from \pstar~by an interval of at least 0.03.
A similar situation prevails in \Eu~(Eu-LSCO), a closely related material, where
$S < 0$ at $p = 0.16$,
$S > 0$ at $p = 0.21$, while \pstar~$\simeq 0.24$ [\onlinecite{Laliberte2011}].

\subsection{Fermi-surface transformation}

Having established that \pstar~is purely the critical point of the pseudogap phase, devoid of superconductivity
or CDW order, let us see what its intrinsic properties are, as may be deduced from transport measurements.
The key signature is a drop in carrier density from $n = 1 + p$ at $p >$~\pstar~to $n = p$
at $p <$~\pstar.
This conclusion can only be reached by looking at both \RH~and $\rho$.
While the drop in \nH~by itself does suggest a drop in $n$, it is not conclusive,
since it could be just a change of Fermi-surface curvature, or a deformation
at roughly constant volume.
It is the huge upturn in $\rho(T)$ that really shows there is a loss of carrier density.

The fact that \nrho~$\simeq p$ at $p = 0.20$, 0.21 and 0.22 is striking (Fig.~8).
The same finding was reported for LSCO, where \nrho~$\simeq p$ at $p = 0.14 - 0.15$ [\onlinecite{Laliberte2016}].
Our data therefore confirm the conclusion of Ref.~\onlinecite{Laliberte2016} that 
the fundamental mechanism for what has been called a ``metal-to-insulator crossover" 
for two decades \cite{Ando1995,Boebinger1996} is a metal-to-metal transition at $T=0$ that
transforms the Fermi surface and
cuts the carrier density down by 1.0 hole per Cu atom.

%%%%%%%%%%%%%%%%%%%%%%%%%%%%%%%%%  Figure 11   %%%%%%%%%%%%%%%%%%%%%%%%%%%%%%%%

\begin{figure}[t]
\centering
\includegraphics[width=8.4cm]{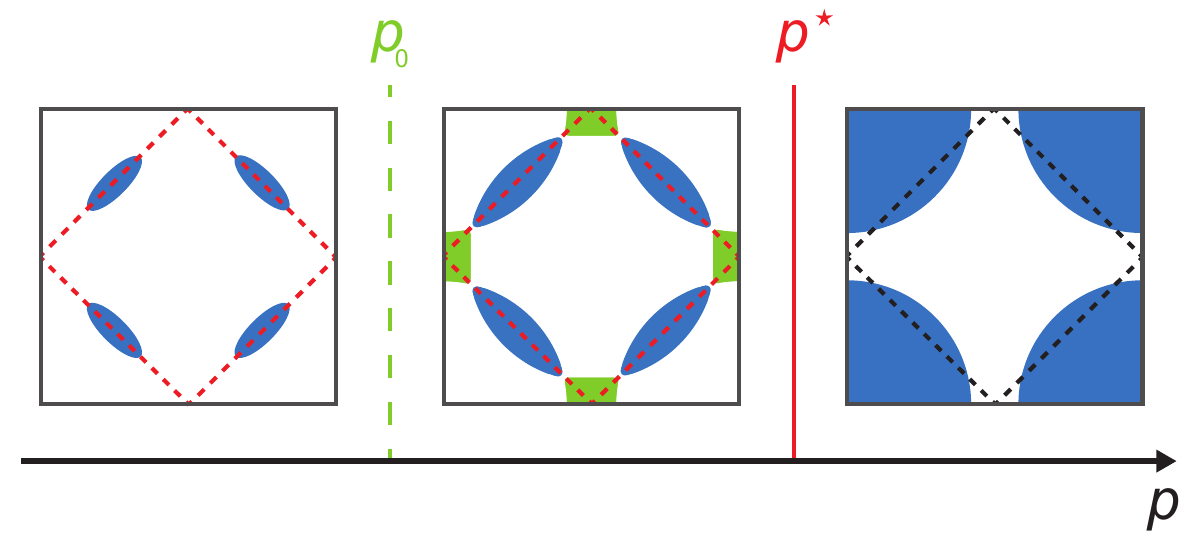}
\caption{
Sketch of the Fermi surface of a single-layer tetragonal cuprate as it evolves with doping
in a scenario where a phase with modulations of wave vector $Q = (\pi, \pi)$ sets in below \pstar.
Above \pstar, the large holelike Fermi surface centered at $(\pi, \pi)$ contains $1+p$ holes (per Cu atom).
Just below \pstar, the new periodicity and associated Brillouin zone (red dashed line)
cause a folding of the large Fermi surface
that produces nodal hole pockets (blue) and anti-nodal electron pockets (green).
With further decrease in $p$, as the modulations and associated gap 
get stronger, the electron pockets shrink and eventually vanish,
below some doping $p_0$. Below $p_0$, the four nodal hole pockets contain
a total of $p$ holes.
}
\label{density}
\end{figure}

%%%%%%%%%%%%%%%%%%%%%%%%%%%%%%%%%%%%%%%%%%%%%%%%%%%%%%%%%%%%%%%%%%%%%%%

One of our important findings is that the pseudogap phase onsets rapidly at $T=0$:
\nrho~drops from $1+p$ to $p$ in a doping interval of at most 0.015 
(Fig.~8), \ie~6~\% of \pstar.
This argues in favour of a transition,
as opposed to a crossover.

The transition in \nH~is wider than in \nrho, and it has additional structure (Fig.~8).
Going back to the raw data of Fig.~5, we see that, at $p = 0.23$,~\RH$(T)$ does not show any upturn at low $T$, while $\rho(T)$ does (Fig.~4).
%
%In fact, \RH$(T)$ at $p = 0.23$ decreases with decreasing $T$, starting with a value 
%at $T = 80$~K that is larger than the value at $p = 0.24$, and ending with a value
%at $T = 10$~K that is the same (Fig.~5).
%
In other words, the drop in \nH~is not detected at $p = 0.23$, but only at $p = 0.22$,
while the drop in \nrho~is clearly seen at $p = 0.23$ (Fig.~8).
%
%In Fig.~8b, we track the discrepancy between \nH~and \nrho~by plotting 
%the ratio \nH/\nrho~vs $p$. 
%
%Equal to unity at $p = 0.24$, the ratio peaks at $p = 0.22$,
%with a value of 2.5, and then it goes back down to $\simeq 1.0$ at $p = 0.20$.

A possible explanation for this difference is the presence
of electronlike carriers in the
Fermi surface of the pseudogap phase,
within a small doping interval immediately below \pstar.
This is why the upturn in \RH$(T)$ (Fig.~5) is less pronounced than it is in $\rho(T)$ (Fig.~2):
electronlike carriers make a negative contribution to \RH~that reduces 
the large (positive) rise due to the loss of 1.0 hole per Cu. high-field Seebeck data were used
A good way to visualize the negative contribution to the Hall response 
made by electronlike carriers is to plot \RH$/ \rho$ vs $T$, as done in Fig.~10(a).
In Nd-LSCO at $p = 0.24$, \RH$/ \rho$ increases monotonically with decreasing $T$ all the way to $T \simeq 0$.
At $p = 0.22$,
%}, 
\RH$/ \rho$ shows the same monotonic increase down to \Tstar, but then it drops below $\sim 30$~K.
This drop relative to monotonic background can only come from a negative contribution to the Hall signal, 
since the mobility keeps increasing monotonically all the way, 
as established by the MR [Fig.~9(b)].

We propose that the narrow peak in \nH~observed in LSCO just below \pstar~$\simeq 0.18$
 (Ref.~\onlinecite{Balakirev2009}) has the same origin.
In Fig.~10(b), we plot \RH$/ \rho$ vs $T$ for LSCO at $p = 0.17 <$~\pstar~and $p = 0.19 >$~\pstar, 
using published data for \RH($T$) 
 (Ref.~\onlinecite{Balakirev2009}) and $\rho(T)$ \cite{Boebinger1996,Cooper2009}.
 We observe the same behavior that we saw in Nd-LSCO [Fig.~10(a)].
 The fact that \RH$(T)$ in LSCO at $p = 0.17$ shows not a reduced rise at low $T$
 (as in Nd-LSCO at $p = 0.22$) but an actual {\it decrease} reinforces the case for an electronlike (negative) contribution
 to the Hall signal.
In the next section, we give a simple example of how electronlike carriers can appear 
as a result of Fermi-surface transformation.

\subsection{Scenario of an antiferromagnetic QCP}

The simplest scenario to explain a transition from $n = 1+p$ to $n = p$ is a
quantum phase transition into a phase of antiferromagnetic (AF) order below 
a QCP at \pstar, with a wave vector $Q = (\pi, \pi)$.
The new periodicity imposed by the spin modulation breaks the translational symmetry
and hence imposes a new, smaller Brillouin zone, sketched by the
dashed line in Fig.~11.
This new zone causes a folding of the original large holelike Fermi surface,
which gets reconstructed into small hole pockets at the ``nodal" positions 
and small electron pockets at the ``anti-nodal" positions.
As the AF moment and associated gap increase with decreasing $p$,
the electron pockets shrink and eventually vanish, below some doping $p_0$,
leaving only the nodal hole pockets (Fig.~11).
By the Luttinger rule, the large Fermi surface above \pstar~contains $1+p$ holes and
the total volume of the four identical nodal hole pockets below $p_0$ 
must be such that $n = p$.

Recently, Storey calculated the Hall coefficient of a typical cuprate as a function of doping
within such an AF scenario [\onlinecite{Storey2016}]. 
As shown in Fig.~12, the value of \RH~at $T=0$ he obtains yields a Hall number
\nH~vs $p$~in good agreement with the YBCO data.
The width of the intermediate regime where electron pockets are present is determined
by how fast the AF gap rises as $p$ decreases below \pstar. 
If the gap grows from zero at \pstar, there will necessarily be an initial regime
containing anti-nodal electron pockets, whose width is controlled by how fast the gap rises. 
For the parameters chosen in the calculation (where the gap at $p=0$ is $J$), 
the width of the intermediate regime between \pstar~and $p_0$ is 0.03,
in good agreement with the observed width in the drop of \nH~vs $p$ for both
YBCO and Nd-LSCO (Fig.~7).

%%%%%%%%%%%%%%%%%%%%%%%%%%%%%%%%%  Figure 12  %%%%%%%%%%%%%%%%%%%%%%%%%%%%%%%%

\begin{figure}[t]
\centering
\includegraphics[width=8.2cm]{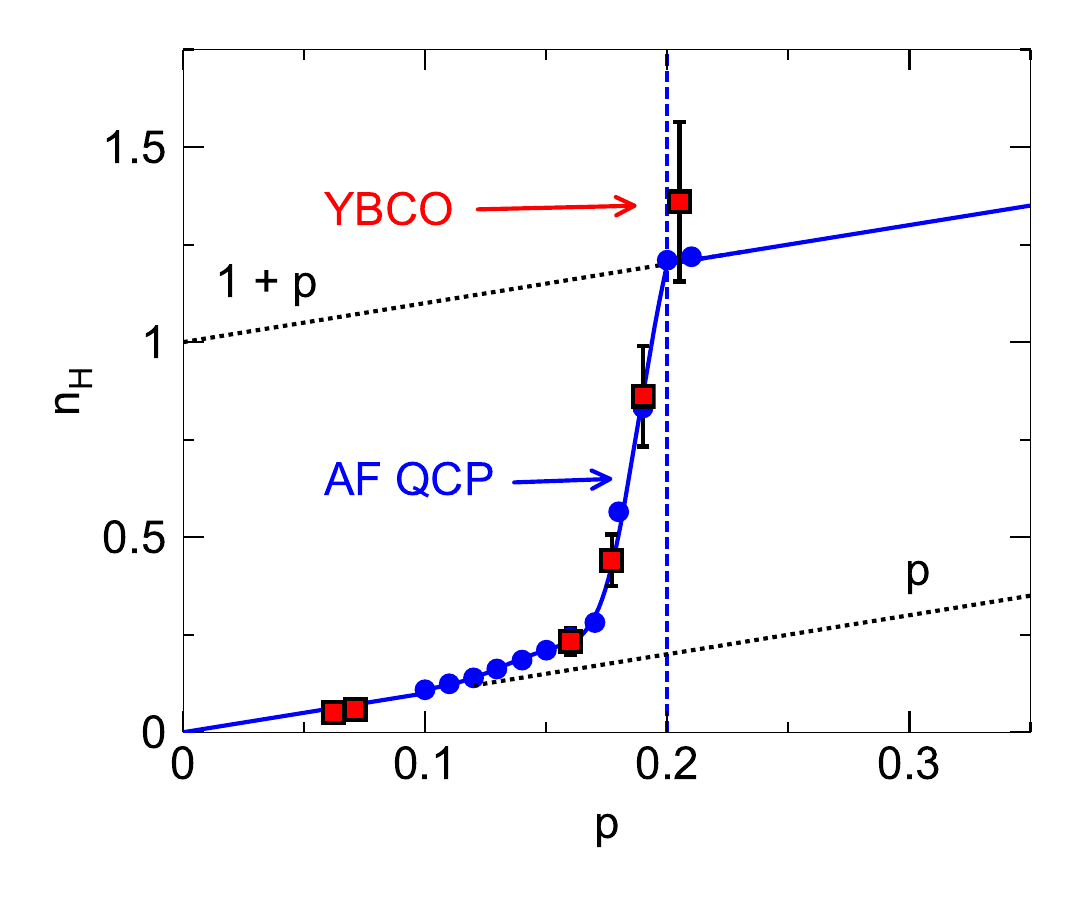}
\caption{
Calculated Hall number \nH~in the $T=0$ limit as a function of doping, 
across a quantum critical point for the onset of antiferromagnetic order,
below \pstar~$=0.20$ (vertical dashed blue line),
with wave vector $Q = (\pi, \pi)$ (blue dots; Ref.~\onlinecite{Storey2016}).
Normal-state \nH~measured in YBCO (red squares),
above $p = 0.15$ (Ref.~\onlinecite{Badoux2016}) 
and below $p = 0.09$ (Ref.~\onlinecite{Segawa2004}),
on either side of the CDW phase.
The solid blue line is a guide to the eye. 
The two dotted lines mark \nH~$=1+p$ and \nH~$=p$, as indicated.
}
\label{Theory-YBCO}
\end{figure}

%%%%%%%%%%%%%%%%%%%%%%%%%%%%%%%%%%%%%%%%%%%%%%%%%%%%%%%%%%%%%%%%%%%%%%%

In summary, the AF scenario
accounts naturally for the observation that $n = p$ 
below \pstar, and it explains why there is a width to the drop in \nH, due to the presence of electron pockets.
It is not clear, however, that such a scenario really applies to hole-doped cuprates.
(Note that it is quite reasonable for electron-doped cuprates [\onlinecite{Armitage2010}].)
In order to confirm its applicability, one would need to detect AF modulations
in the normal state at $T \to 0$.
In Nd-LSCO, magnetic Bragg peaks are observed by neutron diffraction
up to $p = 0.20$, the highest doping investigated so far [\onlinecite{Tranquada1997}],
and the onset temperature \Tsdw~does extrapolate linearly to zero at $p = $~\pstar.
The integrated intensity of the magnetic scattering (proportional to the square of the 
magnetic moment) also extrapolates linearly to \pstar[\onlinecite{Tranquada1997}].
However, the SDW wave vector is $Q = (\pi + \epsilon, \pi)$, not quite $(\pi, \pi)$. 
Whether the incommensurability
would change significantly the resulting carrier density remains to be calculated [\onlinecite{Eberlein2016}].
Also, the magnetism may not be fully static, even at $T=0$, as no magnetic moment is detected
in Nd-LSCO at $p = 0.20$ by muon spin relaxation, a slower probe than neutrons [\onlinecite{Nachumi1998}].
At any rate, slow antiferromagnetic correlations do appear below \pstar~in Nd-LSCO.
The question is whether these cause the Fermi-surface transformation we detect
clearly at \pstar, or whether they are a consequence of it, much as the CDW order 
appears to be a secondary instability of the pseudogap phase \cite{Cyr-Choiniere2017}.

In LSCO, SDW order is observed at low $T$ up to a critical doping $p_{\rm SDW} \simeq 0.13$, in zero magnetic field [\onlinecite{Chang2008a}].
Application of a field moves $p_{\rm SDW}$ up to ~$\sim 0.15$ in $H = 15$~T [\onlinecite{Chang2008a}].
It is conceivable that a field of 60~T, large enough to fully suppress superconductivity in LSCO,
would move $p_{\rm SDW}$ up to \pstar~$= 0.18$, making the phase diagram of LSCO in high field 
qualitatively similar to that of Nd-LSCO in zero field.
(In YBCO, the field needed to suppress superconductivity is 150~T.)

\subsection{Other scenarios}

A number of theoretical scenarios have been proposed to account for the pseudogap phase of cuprate superconductors.
In some, the pseudogap phase is a state that breaks a symmetry.
For example, 
$d$-density-wave order breaks translational symmetry 
with the same $Q$ vector as the commensurate AF state, and therefore
produces the same reduced Brillouin
zone, Fermi-surface pockets (as in Fig.~11), and associated loss of 1.0 hole per Cu atom [\onlinecite{Chakravarty2002}].
Calculations for this state show that the Hall number drops sharply at \pstar [\onlinecite{Chakravarty2001}].

Scenarios without broken translational symmetry could also apply.
In the Yang, Rice, and Zang (YRZ) model [\onlinecite{Yang2006}], 
umklapp scattering derived from the Mott insulator, occurring along a line in 
$k$~space that coincides with the AF Brillouin zone, causes a transformation of the Fermi surface that 
results in small nodal hole pockets with $n = p$, but these are now  confined to one side of the 
umklapp line/AF zone boundary.
There is also a regime of small coexisting antinodal electron pockets immediately below \pstar.
Calculations of \nH~vs $p$ in the YRZ model yield good agreement with experimental data [\onlinecite{Storey2016}].

In the FL* model [\onlinecite{FL*}] and in DMFT solutions to the Hubbard model [\onlinecite{Sakai2009}] ,
small nodal hole pockets
also appear without broken symmetry, but their location is not pinned to the AF zone boundary.

We propose three avenues of investigation that could help discriminate between
the various scenarios.
First, it is important to understand what controls the actual location of the critical point,
which varies considerably amongst hole-doped cuprates [\onlinecite{Cyr-Choiniere2017}],
 \eg~\pstar~$=0.18$~in LSCO \cite{Laliberte2016}
vs~\pstar~$=0.23$~in Nd-LSCO.
Second, the critical point is characterized by two fundamental properties,
both of which should be explained within a single model: 
the drop of carrier density below \pstar, discussed here, and
the linear $T$ dependence
of $\rho(T)$ as $T \to 0$ at \pstar, established in 
LSCO \cite{Cooper2009} and Nd-LSCO [\onlinecite{Daou2009}].
Third, a mechanism for the transformation of the Fermi surface that would account
for the large drop in carrier density below \pstar~should also account for the lack of change
in the mobility, in the regime of disorder scattering at $T=0$.

%%%%%%%%%%%%%%%%%%       SUMMARY  &  OUTLOOK  

\section{Summary}

%%%%%%%%%%%%%%%%%%     Summary  

In summary, 
we performed high-field measurements of the resistivity and Hall coefficient 
in Nd-LSCO across the critical doping where the pseudogap phase ends, at \pstar~$=0.23$.
At $p >$~\pstar, 
\RH$(T$) is flat and it yields a Hall number 
\nH~$\simeq 1 + p$, consistent with a carrier density of $n = 1 + p$ holes per Cu atom.
The resistivity is linear in $T$ as $T \to 0$.
At $p <$~\pstar, 
both $\rho(T)$ and \RH$(T$) exhibit an upturn at low $T$, 
showing that the pseudogap phase causes a drop in carrier density.
Quantitatively, we observe a drop from $n \simeq 1 + p$ at $p >$~\pstar~to
$n \simeq p$ at $p <$~\pstar.
As observed in LSCO [\onlinecite{Laliberte2016}], 
the resistivity of Nd-LSCO reflects the full effect of this loss of carriers,
rising to a value at $T=0$ that is enhanced by a factor $(1+p)/p$ relative to what it would be without
pseudogap.
This implies that the mobility at $T=0$ is essentially unaffected by the opening of the pseudogap
below \pstar, in agreement with the fact that the relative magnetoresistance has the same magnitude on 
both sides of \pstar.

At $T=0$, the change from a metal with $n = 1 + p$ carriers to a metal with $n = p$ carriers
happens very rapidly, within an interval $\delta p/$\pstar~$<Ê6$~\%.
We conclude that the onset of the pseudogap phase at $T=0$ is a transition (vs doping), 
whereas it appears to be a crossover as a function of temperature.

Below \pstar,
we find that the (positive) Hall angle drops at low temperature,
possible evidence for the presence of electronlike carriers in the pseudogapped Fermi surface.
This could explain the small anomalous peak in \nH~vs $p$ 
observed in LSCO \cite{Balakirev2009} and Bi-2201 \cite{Balakirev2003}
just below \pstar.
%

%%%%%%%%%%%%%%%%%%     Outlook  

Our data are quantitatively consistent with the drop in carrier density observed in YBCO
from high-field Hall data [\onlinecite{Badoux2016}] 
and in LSCO from high-field resistivity data [\onlinecite{Laliberte2016}],
both in the size of the drop (by 1.0 hole per Cu) and in the width of the transition ($\delta p \simeq 0.03 - 0.04$).
The Fermi surface transformation across \pstar~observed in all three cuprates 
can be described nicely by a quantum phase transition into a phase 
of long-range AF order with wave vector $(\pi,\pi)$.
However, because there is no evidence of long-range commensurate AF order
at high doping in hole-doped cuprates, the real mechanism may be different.
A key question is whether translational symmetry is broken or not,
and if so, on what length scale.

%%%%%%%%%%%%%%%%%%%%%%%%%%%%%%%%%  Table I  %%%%%%%%%%%%%%%%%%%%%%%%%%%%%%%%

\begin{table}[b]
%\begin{table}[h!]
\begin{center}
\begin{tabular}{ccccc}
\hline
\hline
$x$ & $p$ from \Tc  &	$p$ from \nH & $p$ &  label \\
\hline
0.20     &	   --	                                                      &    0.201 $\pm$ 0.002     &    0.20 $\pm$ 0.002     & 0.20\\
0.21     &      --                                                         &    0.209 $\pm$ 0.003     &    0.21 $\pm$ 0.003     & 0.21 \\
0.22     &     0.220 $\pm$ 0.002      &    0.221 $\pm$ 0.004    &     0.22 $\pm$ 0.003 & 0.22 \\
0.23     &     0.231 $\pm$ 0.002      &    0.230 $\pm$ 0.005     &     0.23 $\pm$ 0.003    & 0.23 \\
0.25    &	 0.236 $\pm$ 0.002       &    0.246 $\pm$ 0.006     &     0.24 $\pm$ 0.005 & 0.24 \\
\hline
\hline
\end{tabular}
\end{center}
\caption{
Estimate of the doping $p$ for each of our five samples of Nd-LSCO (fourth column), 
assuming that on average it is given by their Sr content $x$ (first column).
The $x$ dependence of \Tc~[Fig.~13(a)] and of \nH~[Fig.~13(b)] reveal only small deviations from the relation $p=x$
for the first four samples, corresponding to values and error bars listed in the second and third columns, respectively.
For the fifth sample, with $x=0.25$, the deviation is significant and it points to a doping $p \simeq 0.24$.
}
\end{table}

%%%%%%%%%%%%%%%%%%%%%%%%%%%%%%%%%%%%%%%%%%%%%%%%%%%%%%%%%%%%%%%%%%%%%%%

%%%%%%%%%%%%%%%%%%%%%%%%%%%%%%%%%  Figure 13  %%%%%%%%%%%%%%%%%%%%%%%%%%%%%%%%

\begin{figure}[t]
%\begin{figure}[h!]
\centering
\includegraphics[width=8.2cm]{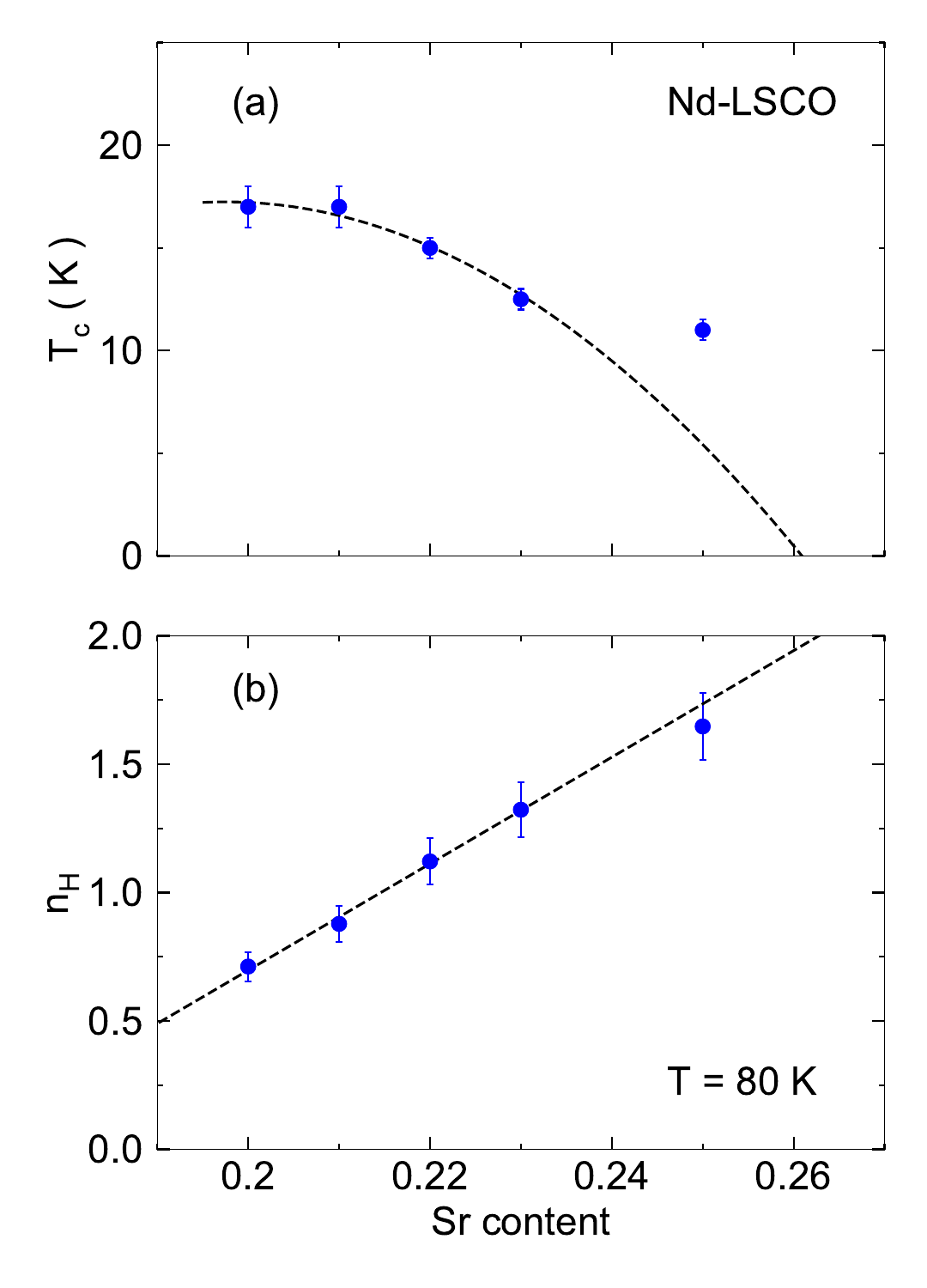}
\caption{
%\textcolor{red}{
(a)
Bulk critical temperature \Tc~of our five samples of Nd-LSCO, measured with a vibrating sample magnetometer,
plotted as a function of Sr content $x$.
The dashed line is a polynomial fit to the four lowest dopings.
(b)
Hall number \nH~of the same samples, measured at $T=80$~K, as a function of $x$.
The dashed line is a linear fit to the four lowest dopings.
%}
}
\label{determine-p}
\end{figure}

%%%%%%%%%%%%%%%%%%%%%%%%%%%%%%%%%%%%%%%%%%%%%%%%%%%%%%%%%%%%%%%%%%%%%%%

%%%%%%%%%%%%%%%%%%%%%%%%%%%%%%%%%  Figure 14  %%%%%%%%%%%%%%%%%%%%%%%%%%%%%%%%

%\begin{figure}[t]
%\begin{figure}[!h]
%\centering
%\includegraphics[width=7.5cm]{figure14-v1.pdf}
%\caption{
%\textcolor{red}{
%Temperature sweeps at $H = 16 $ (blue line) and 33 T (dark blue line) for three of our samples : $p=0.21$ (panel a), 0.22 (panel b) and 0.23 (panel c).
%The red dots are the back extrapolated values extracted from $H^2$ fits on the field sweeps as shown figure \ref{H-isotherms}.
%The black dashed line is a linear fit from high temperature data.
%The red dash line is a linear fit from the red dots and is used to extract the magnetoresistance free value of $\rho (0)$.
%}
%}
%\label{MR}
%\end{figure}

%%%%%%%%%%%%%%%%%%%%%%%%%%%%%%%%%%%%%%%%%%%%%%%%%%%%%%%%%%%%%%%%%%%%%%%

%%%%%%%%%%%%%%%%%      ACKNOWLEDGEMENTS   

\section{ACKNOWLEDGEMENTS}

We thank K. Behnia, M. Charlebois, S. Chatterjee, A. Chubukov, A. Eberlein, M. Ferrero, A. Georges, N. Hussey, S. Kivelson, D. LeBoeuf., S. Lederer, W. Metzner, A. Moutenet, C. Proust, B. Ramshaw, S. Sachdev, A. Sacuto, J. Storey, J. Tranquada, A.-M. Tremblay, and S. Verret for stimulating discussions, and S. Fortier for his assistance with the experiments.
This work was supported by a 
Canada Research Chair,
the Canadian Institute for Advanced Research (CIFAR), 
the National Science and Engineering Research Council of Canada (NSERC), 
the Fonds de recherche du Qu\'ebec - Nature et Technologies (FRQNT), 
and the Canada Foundation for Innovation (CFI).
This work was supported by HFML-RU/FOM, a member of the European Magnetic Field Laboratory (EMFL). J.-S.Z. was supported by the DOD-ARMY grant (W911NF-16-1-0559) in USA.

\appendix

\section{Doping values}

As is usual, we assume that the doping $p$ of Nd-LSCO samples is given by their Sr content $x$, 
\ie~$p = x$. 
Of course, the distribution of Sr atoms in a particular sample depends on the growth conditions 
and a $p$ value slightly away from $x$ is not unusual. 
%
%What matters in our study is not so much the absolute value of $p$, 
%but the doping value of one sample relative to the others. 
%
%(We also want to be able to compare our transport data to ARPES data on samples with the same doping. 
%As it happens, the only ARPES study on Nd-LSCO (Ref.~\onlinecite{Matt2015})
%was performed on crystals from the same batch as our own crystals,
%with $p = 0.20$ ($x = 0.20$) and $p = 0.24$ ($x = 0.25$). 
%So the ARPES-transport comparison is reliable.)
%
We refine the relative dopings of our five samples as follows.
In Fig.~13(a), we plot \Tc~vs $x$. 
We observe that \Tc~is a smoothly-decreasing function of $x$ except for the $x = 0.25$ sample, 
whose \Tc~is much too high (as also found in Ref.~\onlinecite{Daou2009}). 
Its \Tc~value is instead consistent with a doping $p = 0.236$. 
We also observe that \nH~at $T = 80$~K (above \Tstar) increases linearly with $x$ for the first four samples [Fig.~13(b)], 
while the 5th sample (with $x = 0.25$) has a slightly too low value. 
Its \nH~value is consistent with $p = 0.246$. 
We therefore find that $p = x$ within $\pm 0.003$ for the first four samples, 
and $p = 0.24 \pm 0.005$ for the fifth sample (Table I).

\bibliographystyle{apsrev4-1}
\bibliography{Bib}

\end{document}